\documentclass[a4paper,11pt]{article}
\pdfoutput=1 

\usepackage{jcappub} 

\usepackage[T1]{fontenc} 

\usepackage{amsmath}

\allowdisplaybreaks

\title{\boldmath Massive Scalar Field Perturbations of Black Holes Immersed in Chaplygin-Like Dark Fluid}

\author[a]{Ram\'on B\'ecar,}
\author[b]{P. A. Gonz\'alez,}
\author[c]{Eleftherios Papantonopoulos}
\author[d,1]{and Yerko V\'asquez \note{Corresponding author}}

\affiliation[a]{Departamento de Ciencias Matem\'{a}ticas y F\'{\i}sicas, Universidad Catolica de Temuco}
\affiliation[b]{Facultad de
Ingenier\'{i}a y Ciencias, Universidad Diego Portales, Avenida Ej\'{e}rcito
Libertador 441, Casilla 298-V, Santiago, Chile}
\affiliation[c]{Physics Division, School of Applied Mathematical and Physical Sciences, National Technical University of Athens, 15780 Zografou Campus,
    Athens, Greece}

\affiliation[d]{Departamento de F\'isica, Facultad de Ciencias, Universidad de La Serena,\\
Avenida Cisternas 1200, La Serena, Chile}
\emailAdd{rbecar@uct.cl}
\emailAdd{pablo.gonzalez@udp.cl}
\emailAdd{lpapa@central.ntua.gr}
\emailAdd{yvasquez@userena.cl}

\abstract{We consider massive scalar field perturbations in the background of black holes immersed in Chaplygin-like dark fluid (CDF), and we analyze the photon sphere modes, the  de Sitter modes as well as the near extremal modes  and discuss  their dominance, by using the  pseudospectral Chebyshev method and the third order Wentzel-Kramers-Brillouin approximation.  We also discuss the impact of the parameter representing the intensity of the CDF on the families of quasinormal modes. Mainly, we find that the propagation of a massive scalar field is stable in this background, and it is characterized by quasinormal frequencies with a smaller oscillation frequency and a longer decay time  compared to the propagation of the same massive scalar field within the Schwarzschild-de Sitter background.}

\begin{document}
\maketitle
\flushbottom

\section{Introduction}

Recent astrophysical observations revealed that our Universe is currently undergoing accelerated expansion \cite{SupernovaCosmologyProject:1998vns,SupernovaSearchTeam:1998fmf,SupernovaSearchTeam:1998cav}.
This accelerated expansion in General Relativity (GR) can be explained by the presence of a new energy density component with negative pressure and positive energy density, called dark energy (DE). The negative pressure could be originated from the presence of  barotropic perfect fluid which corresponds to the dark energy or to the presence of a cosmological constant. A  relation between the pressure $p$ and the energy density $\rho$, defines the  state equation  $w=p/\rho$ and the recent observations leads to a narrow strip around $w =-1$  of  the equation of state \cite{WMAP:2010qai,Alam:2003fg}. The value  $w =-1$  corresponds to a cosmological constant $\Lambda$ while $w <-1$ is allowed \cite{Caldwell:1999ew,Caldwell:2003vq} which indicates the presence of a phantom field with negative kinetic energy.

On the other hand,  the large scale structure distributions in the whole Universe and the missing mass in individual galaxies leads to the existence of a new form of matter, termed dark matter (DM), which is assumed to have negligible pressure. The introduction of dark matter may explain the discrepancy between the predicted rotation curves of galaxies when only including luminous matter and the actual (observed) rotation  curves which differ significantly \cite{deAlmeida:2018kwq,Harada:2022edl,Shabani:2022buw}. Besides that, in a effort to explain the dark components of the Universe and  to understand and
describe the thermal history of the Universe in a unified way novel models that combine dark matter and dark energy were proposed. Among these unified dark fluid models, the Chaplygin gas and its related generalizations have gained significant attention in elucidating the observed accelerated expansion of the Universe~\cite{Kamenshchik:2001cp,Bilic:2001cg,Bento:2002ps}.

The main motivation of considering  the Chaplygin gas \cite{Bento:2002ps} was to modify the equation of state in such a way as to
mimic a pressureless fluid at the early stages of the evolution of the Universe, and a quintessence scalar field sector at late times,
which tends asymptotically to a cosmological constant. Recently the Chaplygin gas has been used to address the Hubble tension problem \cite{Sengupta:2023yxh} and studying the growth of cosmological perturbations~\cite{Abdullah:2021tee}. In \cite{Li:2019lhr} a charged static spherically-symmetric black hole surrounded by Chaplygin-like dark fluid (CDF)  with the equation of state $p=-\frac{B}{\rho}$ in the Lovelock gravity theory was considered, and an analytical solution was found and also the thermodynamics was  studied.

One of the most well studied compact object is the black hole (BH). Black holes in our Universe may be affected by astronomical
environments, such as the DM near the BHs \cite{Ferrer}-\cite{Xu}. It is believed that $90\%$   of galaxies  are composed of DM \cite{Jusufi}.
A BH embedded in the center of a galactic dark matter halo profile was studied in \cite{Cardoso:2021wlq} and axial and polar perturbations of such configurations were calculated in \cite{Cardoso:2022whc} while gravitational-wave imprints of such  compact oblects embedded in  galactic-scale environments were studied in \cite{Destounis:2022obl}.
The electromagnetic field does not interact with DM and therefore the propagation of light is possible inside the dark matter halo. There are many  studies to learn whether the BH shadow could be affected by the tidal forces induced by the invisible matter \cite{Hou:2018avu,Haroon:2018ryd,Hou:2018bar}.
However, due to a particular equation of state for the dark matter which is considered in most of the studies, the results  look highly model-dependent.
The most reliable way to detect DM outside the BH is to study the gravitational wave signal, produced during the ringdown phase
generating a series of echoes. Various models of black holes surrounded by dark matter have been studied \cite{Jusufi:2022jxu}-\cite{Xavier:2023exm} and their stability in astrophysical environments was studied in \cite{Destounis:2023ruj}.

In a recent paper \cite{Li:2024abk} the shadows and optical appearances of a black hole immersed in   CDF   with thin disk accretion and spherically symmetric accretion were studied.  The geodesic structure, shadow, and optical appearance of such a black hole was investigated. They found that the shadow and optical appearance of the CDF black hole  imposed constraints on the CDF model in the Universe from the observation of Event Horizon Telescope. Assuming that  CDF is causing cosmic accelerated expansion for a static observer located far from both event and cosmological horizons using the observational data on shadow radius they constrained the parameters in the CDF model. For a static observer located near the pseudo-cosmological horizon, they  obtained  a black hole optical image with impact parameter as a coordinate scale.

The aim of this work is to study the propagation of massive scalar fields in a BH spacetime immersed in CDF and analyze the photon sphere modes as well as the  de Sitter modes and  by using the  pseudospectral Chebyshev method and the third order Wentzel-Kramers-Brillouin approximation to discuss  their dominance.
We analytically  discuss the impact of the parametric space representing the intensity of the CDF on both families of quasinormal modes (QNMs). The QNMs give an infinite discrete spectrum which consists of complex frequencies, $\omega = \omega_R + i\omega_I$, where the real part $\omega_R$ determines the oscillation timescale of the modes, while the complex part $\omega_I$ determines their exponential decaying timescale. In the case that  the background is the Schwarzschild and the Kerr BHs it was found that for gravitational perturbations the longest-lived modes are  the ones with lower angular number $\ell$.  This behaviour can be understood from the fact that the more energetic modes with high angular number $\ell$ would have faster
decaying rates.

However, if the perturbed scalar field is massive, then the longest-lived modes are the ones with higher angular number. This is happening because there is a critical mass of the scalar field  where the behaviour of the decay rate of the QNMs is inverted. This different behaviour  can be obtained from the condition $Im(\omega)_{\ell}=Im(\omega)_{\ell+1}$ in the {\it eikonal} limit, that is when $\ell \rightarrow \infty$. It was found that this anomalous behaviour in the quasinormal frequencies (QNFs) is possible in asymptotically flat, in asymptotically dS and  in asymptotically AdS spacetimes. However,  it was shown  that  the critical mass exist for asymptotically flat and for asymptotically dS spacetimes and it is not present in asymptotically AdS spacetimes for large and intermediate BHs. This behaviour has been extensively studied for scalar fields \cite{Konoplya:2004wg, Konoplya:2006br,Dolan:2007mj, Tattersall:2018nve,Lagos:2020oek, Aragon:2020tvq,Aragon:2020xtm,Fontana:2020syy,Becar:2023jtd, Becar:2023zbl} as well as charged scalar fields \cite{Gonzalez:2022upu,Becar:2022wcj} and fermionic fields \cite{Aragon:2020teq}, in BH spacetimes. The  anomalous decay in accelerating black holes was studied in \cite{Destounis:2020pjk}. Also, it has been recently studied for scalar fields in wormhole spacetimes \cite{Gonzalez:2022ote,Alfaro:2024tdr}. Motivated by this anomalous behaviour of massive scalar field perturbations, in this work we also study the propagation of massive scalar fields  for  higher values of $\ell$ in the background of a BH immersed in CDF and we find the parametric space  in which  there is an anomalous decay rate of QNMs  for the photon sphere modes. Also, we show that this behaviour is present for small values of $\ell$ by using the pseudospectral Chebyshev method.

The work is organized as follows. Section \ref{setup}  provides the theoretical framework and the setup of the theory. In Section \ref{perturbations} we discuss the scalar field perturbations. In Section \ref{WKBJ} we study the photon sphere modes. In Section \ref{desitter} we discuss the de Sitter modes. Then, in Section \ref{nemodes} we study the near extremal modes. In Section \ref{family} we study the dominance family modes and finally in Section \ref{conclusion} are our conclusions.


\section{The setup of the theory}
\label{setup}

In this section we review the work presented in \cite{Li:2024abk}, in which a black hole is immersed in a cosmological Chaplygin-like dark fluid (CDF) which is characterized by the equation of state $p=-B/\rho$ and also the energy density of the fluid is influenced by the parameter $q$,  see Eq. (\ref{CDFenergydensity}). For a spacetime that is static and spherically symmetric, the following metric was considered
\begin{equation}
ds^2=-f(r)dt^2+\frac{1}{g(r)} dr^2+r^2d\Omega^2~,\label{dsf}
\end{equation}
where $f(r)$ and $g(r)$ represent general functions dependent on the radial coordinate $r$. The stress-energy tensor describing a perfect fluid is given by
\begin{equation}
T_{\mu\nu}= (\rho+p) u_\mu u_\nu +pg_{\mu\nu}~, \label{Stress Energy Tensor}
\end{equation}
here $\rho$ and $p$ denote the energy density and isotropic pressure, respectively, as measured by an observer moving with the fluid and $u_\mu$ is the four-velocity of the fluid. However, they proposed that  the CDF is anisotropic and following \cite{Raposo:2018rjn}, they expressed  the  stress-energy tensor  as
\begin{equation}
T_{\mu\nu}= \rho u_\mu u_\nu +p_r k_\mu k_\nu +p_t \Pi_{\mu\nu}~, \label{Stress Energy Tensor}
\end{equation}
where $p_r$ and $p_t$ represent the radial and tangential pressure, respectively, $u_\mu$ is the fluid four-velocity, and $k_\mu$ is a unit spacelike vector orthogonal to $u_\mu$. The vectors $u_\mu$ and $k_\mu$ satisfy $u_\mu u^{\mu}=-1$, $k_\mu k^\mu=1$, and $u^{\mu}k_\mu=0$.  The projection tensor is defined as $\Pi_{\mu\nu} = g_{\mu\nu}+u_\mu u_\nu-k_\mu k_\nu$, projecting onto a two-surface orthogonal to $u^{\mu}$ and $k^{\mu}$. Then they considered that the anisotropic fluid CDF is characterized by a non-linear equation of state, expressed as $p=-\frac{B}{\rho}$, where $B$ is a positive constant. Ensuring $p_r=-\rho$, the tangential pressure is derived as $p_t=\frac{1}{2}\rho-\frac{3B}{2\rho}$.

The gravitational equations were found to be
\begin{equation}\label{graviequations}
\frac{1}{r^2}(f+rf'-1)=-\rho~,\hspace{1cm} \frac{1}{2r}(2f'+rf'')=\frac{1}{2}\rho-\frac{3B}{2\rho}~.
\end{equation}

Using the energy-tensor (\ref{Stress Energy Tensor})  they found that the energy density of CDF is
\begin{equation}
\rho(r)=\sqrt{B+\frac{q^2}{r^6}}~, \label{CDFenergydensity}
\end{equation}
where $q>0$ is a normalization factor representing the intensity of the CDF.  For small radial coordinates (i.e., $r^6\ll q^2/B$), the CDF energy density can be approximated by
\begin{equation}
\rho(r)\approx\frac{q}{r^3}~, \label{CDFenergydensitysmallr}
\end{equation}
indicating that the CDF behaves like a matter content with an energy density varying as $r^{-3}$. For large radial coordinates (i.e., $r^6\gg q^2/B$), the energy density behaves as
\begin{equation}
\rho(r)\approx \sqrt{B}~, \label{CDFenergydensitylarger}
\end{equation}
suggesting that the CDF acts as a positive cosmological constant at a large-scale regime.

Substituting Eq.~(\ref{CDFenergydensity}) into Eq.~(\ref{Stress Energy Tensor}), and using the gravitational equations (\ref{graviequations}) they derived the analytical solution for the metric function $f(r)$,
\begin{equation}
f(r)=1-\frac{2M}{r}-\frac{r^2}{3}\sqrt{B+\frac{q^2}{r^6}}+\frac{q}{3r}{\rm arcsinh}\frac{q}{\sqrt{B}r^3}~. \label{ffr}
\end{equation}
Now, from the equation $f(r_h)=0$, where $r_h$ is the event horizon radius, $M$ can be written in terms of $r_h$, $q$ and $B$. Additionally, using the scaled coordinate $\tilde{r} = \frac{r}{r_h}$, and $\tilde{q} = \frac{q}{r_h}$ and $\tilde{B} = B r_h^4$, the metric function can be written as
\begin{equation}
f(\tilde{r}) = 1 - \frac{1}{\tilde{r}}- \frac{\tilde{r}^2}{3}\sqrt{\tilde{B}+\frac{\tilde{q}^2}{\tilde{r}^6}}+ \frac{1}{3 \tilde{r}}A + \frac{\tilde{q}}{3 \tilde{r}} {\rm arcsinh} \left( \frac{\tilde{q}}{\sqrt{\tilde{B}} \tilde{r}^3} \right)  - \frac{\tilde{q}}{3 \tilde{r}} {\rm arcsinh} \left( \frac{\tilde{q}}{\sqrt{\tilde{B}}}\right)\,.
\end{equation}

As can be seen from the above expressions of the metric function, the parameters $\tilde{B}$ and $\tilde{q}$ play a decisive role in the behaviour of the {CDF}. Note that for $\tilde{q}=0$, the geometry represents the Schwarzschild-de Sitter (dS) spacetime with the cosmological constant given by $\Lambda = \sqrt{\tilde{B}}/r_h^2$. Therefore, as $\tilde{q}$—the intensity of the {CDF}—increases, we will be able to analyze the impact of the {CDF} compared to the Schwarzschild-dS spacetime.

  Also, for small values of the argument  $z = \frac{\tilde{q}}{\sqrt{\tilde{B}}} = \frac{q}{\sqrt{B}r_h^3} <<1$ in the arcsinh function, the metric function can be expanded as 
\begin{equation}\label{appr}
f(\tilde{r}) \approx -\frac{1}{3} \sqrt{\tilde{B}} \tilde{r}^2+1 -\frac{1 -\sqrt{\tilde{B}}/3}{\tilde{r}}-z^2 \frac{\sqrt{\tilde{B}} \left(\tilde{r}^3-1\right)}{6 \tilde{r}^4} + \mathcal{O}(z^4) \,,
\end{equation}
indicating that the term proportional to $z^2$ represents the first-order correction to the Schwarzschild-dS spacetime.

\section{Scalar field perturbations}
\label{perturbations}

The QNMs of scalar perturbations in the background of the metric presented in section \ref{setup}
are given by the scalar field solution of the Klein-Gordon equation
\begin{equation}
\frac{1}{\sqrt{-g}}\partial _{\mu }\left( \sqrt{-g}g^{\mu \nu }\partial_{\nu } \varphi \right) =m^{2}\varphi \,,  \label{KGNM}
\end{equation}%
with suitable boundary conditions for a BH geometry. In the above expression $m$ is the mass
of the scalar field $\varphi $. Now, by means of the following ansatz
\begin{equation}
\varphi =e^{-i\omega t} R(r) Y(\Omega) \,,\label{wave}
\end{equation}%
the Klein-Gordon equation reduces to
\begin{equation}
\frac{1}{r^2}\frac{d}{dr}\left(r^2 f(r)\frac{dR}{dr}\right)+\left(\frac{\omega^2}{f(r)}-\frac{\kappa^{2}}{r^2}-m^{2}\right) R(r)=0\,, \label{radial}
\end{equation}%
where $-\kappa^{2}=-\ell (\ell+1)$ represents the eigenvalues of the Laplacian on the two-sphere and $\ell$ is the multipole number, which can take values $\ell=0,1,2,...$.
Now, defining $R(r)=\frac{F(r)}{r}$
and by using the tortoise coordinate $r^*$ given by
$dr^*=\frac{dr}{f(r)}$,
 the Klein-Gordon equation can be written as a one-dimensional Schr\"{o}dinger-like equation
 \begin{equation}\label{ggg}
 \frac{d^{2}F(r^*)}{dr^{*2}}-V_{eff}(r)F(r^*)=-\omega^{2}F(r^*)\,,
 \end{equation}
 with an effective potential $V_{\text{eff}}(r)$, which  parametrically thought,  $V_{eff}(r^*)$, is given  by
  \begin{equation}\label{pot}
 V_{eff}(r)=\frac{f(r)}{r^2} \left(\kappa^{2} + m^2 r^2+f^\prime(r)r\right)\,.
 \end{equation}
 In terms of dimensionless quantities, Eq.(\ref{ggg}) can be written as
\begin{equation}
 \frac{d^{2}F(\tilde{r}^*)}{d\tilde{r}^{*2}}-\tilde{V}_{eff}(\tilde{r})F(\tilde{r}^*)=-\tilde{\omega}^{2}F(\tilde{r}^*)\,,
 \end{equation}
 where
 \begin{equation}
 \tilde{V}_{eff}(\tilde{r})=\frac{f(\tilde{r})}{\tilde{r}^2} \left(\kappa^{2} + \tilde{m}^2 \tilde{r}^2+f^\prime(\tilde{r})\tilde{r}\right)\,,
 \end{equation}
 where
 $\tilde{\omega} = \omega r_h$, and $\tilde{m}= m r_h$.

\begin{figure}[h!]
\begin{center}
\includegraphics[width=0.4\textwidth]{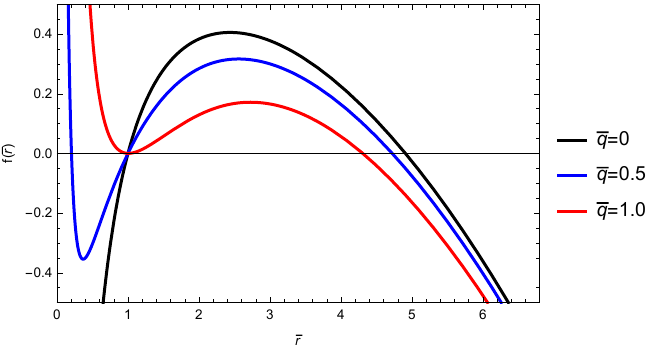}
\includegraphics[width=0.4\textwidth]{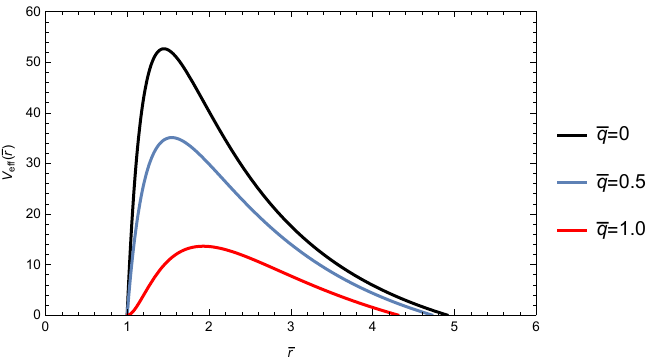}
\end{center}
\caption{The metric function $f(\tilde{r})$ (left panel), and  the effective potential $\tilde{V}_{eff}(\tilde{r})$ (right panel) as a function of $\tilde{r}$, with $\tilde{B}=0.01$. In the left panel $\tilde{r}_{\Lambda}\approx 4.908, 4.723, 4.307$ (pseudo-cosmological horizon), for $\tilde{q}=0, 0.5, 1.0$, respectively. Right panel for $\tilde{m}=0$, and $\ell=20$. }
\label{fV1}
\end{figure}

\begin{figure}[h!]
\begin{center}
\includegraphics[width=0.4\textwidth]{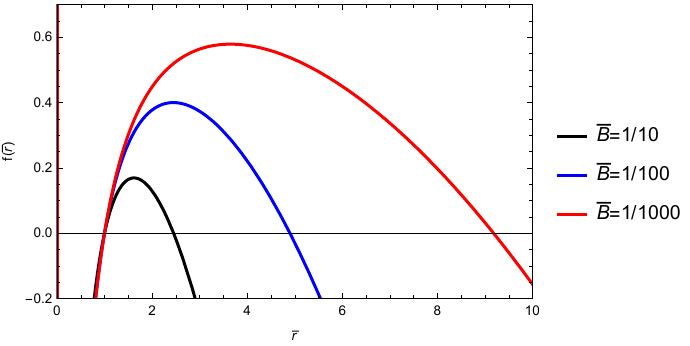}
\includegraphics[width=0.4\textwidth]{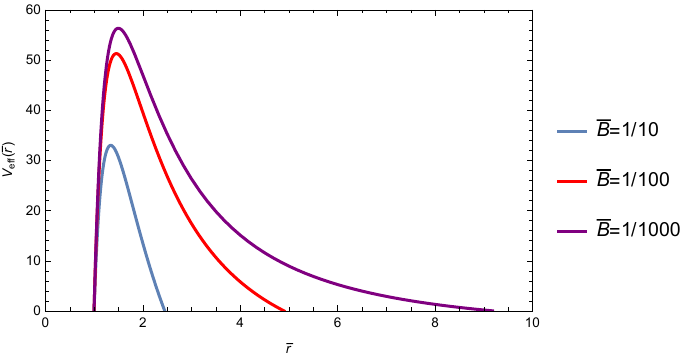}
\end{center}
\caption{The metric function $f(\tilde{r})$ (left panel), and  the effective potential $\tilde{V}_{eff}(\tilde{r})$ (right panel) as a function of $\tilde{r}$, with $\tilde{q}=0.1$. In the left panel $\tilde{r}_{\Lambda}\approx  2.450, 4.897, 9.179$, for $\tilde{B}= 0.1, 0.01, 0.001 $, respectively. Right panel for $\tilde{m}=0$, and $\ell=20$.}
\label{fV2}
\end{figure}

The profiles of the lapse function $f(\tilde{r})$ and the effective potential $\tilde{V}_{eff}(\tilde{r})$ are shown for different values of the parameters $\tilde{q}$ and $\tilde{B}$ in Fig. \ref{fV1} and Fig. \ref{fV2}, respectively.
In the left panel of Fig. \ref{fV1}, it is evident that the position of the pseudo-cosmological horizon decreases with an increasing parameter $\tilde{q}$. This shift, as observed from the potential's perspective in the right panel, leads to a reduction in the communication region between the event horizon $\tilde{r}_h$ and the pseudo-cosmological horizon $\tilde{r}{_\Lambda}$, thereby lowering the potential barrier's height. Conversely, the left panel of Fig. \ref{fV2} demonstrates that an increase in the parameter $\tilde{B}$ corresponds to a decrease in the pseudo-cosmological horizon, accompanied by a simultaneous reduction in the potential barrier's height, as depicted in the right panel.

\section{Photon sphere modes}
\label{WKBJ}

\subsection{Small values of $\ell$}

\begin{figure}[h]
\begin{center}
\includegraphics[width=0.45\textwidth]{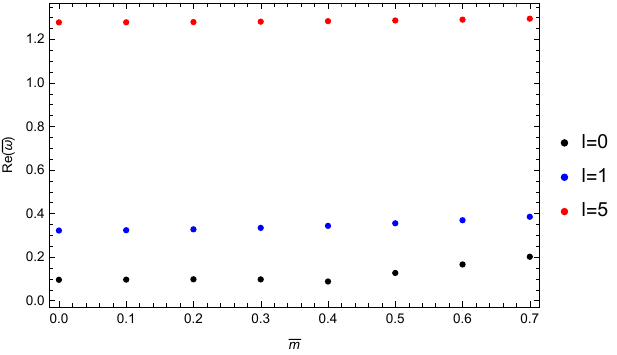}
\includegraphics[width=0.45\textwidth]{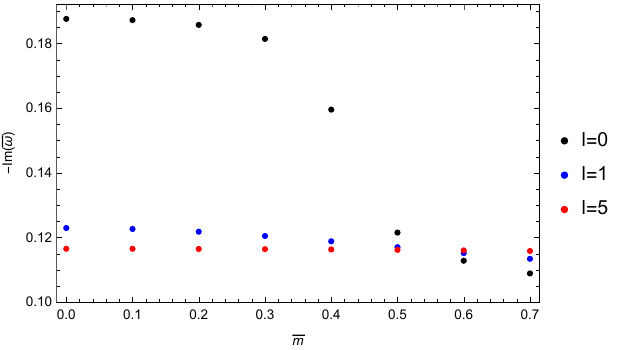}\\
\end{center}
\caption{The behaviour of $Re(\tilde{\omega})$ and $- Im(\tilde{\omega})$ for a massive scalar field as a function of $\tilde{m}$ with $\tilde{q}=0.1$, $\tilde{B}=0.2$, $n=0$ (fundamental frequency), and $\ell = 0, 1, 5$.}
\label{omega1}
\end{figure}

Now, in order to compute the QNFs for small values of the angular number $\ell$, we will solve numerically the differential equation (\ref{radial}) by using the pseudospectral Chebyshev method, for an exhaustive review of the method see for instance \cite{Boyd}. First, it is convenient to perform a change of variable in order to compactify the radial coordinate to the range $[0,1]$. Thus, we define the change of variable $y=(\tilde{r}-1)/(\tilde{r}_{\Lambda}-1)$, where $\tilde{r}_{\Lambda} = r_{\Lambda}/r_h$ is the scaled radius of the pseudo-cosmological horizon. Therefore, now the event horizon is located at $y=0$ and the pseudo cosmological horizon at $y=1$.
The radial equation (\ref{radial}) becomes
\begin{eqnarray} \label{rad}
\notag && f(y) R''(y) + \left( \frac{2 \left(    \tilde{r}_{\Lambda}- 1  \right) f(y)}{1+\left( \tilde{r}_{\Lambda}-1 \right) y } + f'(y) \right) R'(y) \\
&& + \left( \tilde{r}_{\Lambda}-1  \right)^2 \left( \frac{\tilde{\omega}^2}{f(y)}- \frac{ \ell(\ell+1)}{\left( 1 + \left( \tilde{r}_{\Lambda}-1 \right)y \right)^2} -\tilde{m}^2  \right) R(y)=0\,.
\end{eqnarray}
In the vicinity of the horizon ($y \rightarrow 0$) the function $R(y)$ behaves as
\begin{equation}
R(y)=C_1 e^{-\frac{i \tilde{\omega} \left( \tilde{r}_{\Lambda}-1 \right)}{f'(0)} \ln{y}}+C_2 e^{\frac{i \tilde{\omega} \left( \tilde{r}_{\Lambda}-1 \right)}{f'(0)} \ln{y}} \,.
\end{equation}
Here, the first term represents an ingoing wave and the second represents an outgoing wave near the black hole horizon.
So, imposing the requirement of only ingoing waves at the horizon, we fix $C_2=0$. On the other hand, at the cosmological horizon the function $R(y)$ behaves as
\begin{equation}
R(y)= D_1 e^{-\frac{i \tilde{\omega} \left( \tilde{r}_{\Lambda}-1 \right)}{f'(1)} \ln{(1-y)}}+D_2 e^{\frac{i \tilde{\omega} \left( \tilde{r}_{\Lambda}-1 \right)}{f'(1)} \ln{(1-y)}}  \,.
\end{equation}
Here, the first term represents an outgoing wave and the second represents an ingoing wave near the cosmological horizon. So, imposing the requirement of only ingoing waves on the cosmological horizon requires $D_1=0$.
Therefore, taking the behaviour of the scalar field at the event and cosmological horizons we define the following ansatz
\begin{equation}
R(y)= e^{-\frac{i \tilde{\omega} \left( \tilde{r}_{\Lambda}-1 \right)}{f'(0)} \ln{y}} e^{\frac{i \tilde{\omega}  \left( \tilde{r}_{\Lambda}-1 \right)  }{f'(1)} \ln{(1-y)}} F(y) \,.
\end{equation}
Then, by inserting the above ansatz for $R(y)$ in Eq. (\ref{rad}), a differential equation for the function $F(y)$ is obtained. The solution for the function $F(y)$ is assumed to be a finite linear combination of the Chebyshev polynomials, and it is inserted into the differential equation for $F(y)$. Also, the interval $[0,1]$ is discretized at the Chebyshev collocation points. Then, the differential equation is evaluated at each collocation point. So, a system of algebraic equations is obtained, and it corresponds to a generalized eigenvalue problem, which is solved numerically to obtain the QNFs ($\tilde{\omega}$). In appendix \ref{Accuracy} we show the accuracy of the numerical technique used. In tables \ref{TCH1}, and \ref{TCH2} we show some QNFs for a massless scalar field. We can observe that the longest-lived modes are the ones with highest angular number $\ell$. Also, the frequency of oscillation increases when the angular number $\ell$ increases. When the parameter $\tilde{q}$ or $\tilde{B}$ increases the frequency of oscillation and the absolute value of the imaginary part of the QNFs  decrease. However, when the scalar field acquires a mass
an anomalous decay rate is observed, i.e,  for $\tilde{m}<\tilde{m}_c$, the longest-lived modes are the ones with highest angular number $\ell$; whereas, for $\tilde{m}>\tilde{m}_c$, the longest-lived modes are the ones with smallest angular number, as it is shown in Fig. \ref{omega1}.\\

\begin{table}[h]
\centering
\caption{The fundamental $n=0$ QNFs ($\tilde{\omega}$) for massless scalar perturbations with $\tilde{B}=0.1$ for several values of the angular momentum $\ell$, and the parameter $\tilde{q}$. Here, the QNFs are obtained via the pseudospectral Chebyshev method using a number of Chebyshev polynomials in the range $95$-$100$, with eight decimal places of accuracy for the QNFs.
}
\label{TCH1}
\resizebox{1.\textwidth}{!}{
\scriptsize
\begin{tabular}{|c|c|c|c|c|}
\hline
$\ell$ &
$\tilde{q}=0$ &
 $\tilde{q}=0.25$ & $\tilde{q}=0.50$ & $\tilde{q}=0.75$ \\
\hline
0  & $\pm 0.13454127 - 0.20842337  i$ & $\pm 0.12550553 - 0.19140666  i$  & $\pm 0.09938434 - 0.14999504  i$  & $\pm 0.05391235 - 0.09586576 i$  \\
1  &  $\pm 0.39905176 - 0.15146493 i$  & $\pm 0.38013437 - 0.13739895 i$   & $\pm 0.32906769 - 0.10387152 i$  & $\pm 0.24701921 - 0.06249760 i$   \\
2  & $\pm 0.69120033 - 0.14453226 i$  &  $\pm 0.65835302 - 0.13145177 i$  &  $\pm  0.57025969 - 0.10008029 i$ &  $\pm 0.42991692 - 0.06114148 i$ \\
3  & $\pm 0.97773916 - 0.14283854 i$  & $\pm 0.93120165 - 0.13000811 i$  &  $\pm 0.80659973 - 0.09915654 i$ &   $\pm 0.60868509 - 0.06080756 i$ \\
5  & $\pm 1.54619199 - 0.14183109 i$ &  $\pm 1.47252871 - 0.12915284 i$  &  $\pm 1.27546731 - 0.09860783 i$ &  $\pm 0.96305571 - 0.06060745 i$  \\
\hline
\end{tabular}
}
\end{table}

\begin{table}[h]
\centering
\caption{The fundamental $n=0$ QNFs ($\tilde{\omega}$) for massless scalar perturbations with $\tilde{q}=0.1$ for several values of the angular momentum $\ell$, and the parameter $\tilde{B}$. Here, the QNFs are obtained via the pseudospectral Chebyshev method using a number of Chebyshev polynomials in the range $95$-$100$, with eight decimal places of accuracy for the QNFs.}
\label{TCH2}
\scriptsize
\begin{tabular}{|c|c|c|c|}
\hline
$\ell$  & $\tilde{B}=0.20$ & $\tilde{B}=0.40$ & $\tilde{B}=0.60$ \\
\hline
0  & $\pm 0.09557385 - 0.18765413  i$   & $\pm 0.04858808 - 0.14467213  i$  & $\pm 0.02027106 - 0.09754042 i$  \\
1   & $\pm 0.32181504 - 0.12294454 i$   & $\pm 0.21847554 - 0.08312963  i$   & $\pm 0.13703658 - 0.05171468 i$  \\
2    & $\pm 0.56766592 - 0.11827513 i$   & $\pm 0.39191651 - 0.08153609 i$  & $\pm 0.24744404 - 0.05134169 i$  \\
3   & $\pm 0.80620090 - 0.11717777  i$   & $\pm 0.55873852 - 0.08115806  i$  & $\pm 0.35335874 - 0.05125133  i$  \\
5   & $\pm 1.27786185 - 0.11653115  i$   & $\pm 0.88761598 - 0.08093477 i$  & $\pm 0.56191224 - 0.05119769 i$  \\
\hline
\end{tabular}
\end{table}

\newpage

Now, to study the effects of the CDF, we present in Table \ref{TCH2Sch} the QNFs for the parameter $\tilde{q}=0$, representing Schwarzschild-dS black holes with $\Lambda=\sqrt{\tilde{B}}/r_h^2$. By comparing the results in Table \ref{TCH2} with those in Table \ref{TCH2Sch}, it is evident that both the real part of the QNFs and the absolute value of the imaginary part of the QNFs decrease as the parameter $\tilde{q}$ increases from $0$ to $0.1$. This indicates that the photon sphere modes oscillate at lower frequencies and decay more rapidly for black holes immersed in a Chaplygin-like dark fluid.

\begin{table}[h]
\centering
\caption{The fundamental $n=0$ QNFs ($\tilde{\omega}$) for massless scalar perturbations with $\tilde{q}=0$ for several values of the angular momentum $\ell$, and the parameter $\tilde{B}$. Here, the QNFs are obtained via the pseudospectral Chebyshev method using a number of Chebyshev polynomials in the range $95$-$100$, with eight decimal places of accuracy for the QNFs.}
\label{TCH2Sch}
\scriptsize
\begin{tabular}{|c|c|c|c|}
\hline
$\ell$  & $\tilde{B}=0.20$ & $\tilde{B}=0.40$ & $\tilde{B}=0.60$ \\
\hline
0  & $\pm 0.09681708 - 0.19022158 i$   & $\pm 0.04956709 - 0.14710518 i$  & $\pm 0.02100659 - 0.09999739 i$  \\
1  & $\pm 0.32445313 - 0.12487380 i$   & $\pm 0.22094595 - 0.08468397 i$  & $\pm 0.13966025 - 0.05310306 i$  \\
2   & $\pm 0.57241173 - 0.12006287 i$   & $\pm 0.39649319 - 0.08302094 i$  & $\pm 0.25230817 - 0.05270376 i$ \\
3  & $\pm 0.81296833 - 0.11893163 i$   & $\pm 0.56530350 - 0.08262626 i$  & $\pm 0.36034218 - 0.05260693 i$  \\
5  & $\pm 1.28861151 - 0.11826473 i$   & $\pm 0.89808008 - 0.08239303 i$  & $\pm 0.57305076 - 0.05254942 i$  \\
\hline
\end{tabular}
\end{table}


\subsection{High values of $\ell$}

In this section, in order to get some analytical insight of the behaviour of the QNFs up to third order beyond the eikonal limit $\ell \rightarrow \infty$, we use the method based on the Wentzel-Kramers-Brillouin (WKB) approximation initiated by Mashhoon \cite{Mashhoon} and by Schutz and Iyer \cite{Schutz:1985km}. Iyer and Will computed the third order correction \cite{Iyer:1986np}, and then it was extended to the sixth order \cite{Konoplya:2003ii}, and recently up to the 13th order \cite{Matyjasek:2017psv}, see also \cite{Konoplya:2019hlu}.

This method has been used to determine the QNFs for both asymptotically flat and asymptotically de Sitter black holes. This is possible because the WKB method is applicable to effective potentials resembling potential barriers that approach to a constant value at both the horizon and spatial infinity/cosmological horizon \cite{Konoplya:2011qq}.
However,
only the family of QNMs associated to the photon sphere can be obtained with this method. The QNMs are determined by the behaviour of the effective potential near its maximum value $V(r^*_{max})$. The Taylor series expansion of the potential around its maximum is given by
\begin{equation}
V(r^*)= V(r^*_{max})+ \sum_{i=2}^{\infty} \frac{V^{(i)}}{i!} (r^*-r^*_{max})^{i} \,,
\end{equation}
where
\begin{equation}
V^{(i)}= \frac{d^{i}}{d r^{*i}}V(r^*)|_{r^*=r^*_{max}}\,,
\end{equation}
corresponds to the $i$-th derivative of the potential with respect to $r^*$ evaluated at the position of the maximum of the potential $r^*_{max}$. Using the WKB approximation up to third order beyond the eikonal limit, the QNFs are given by the following expression \cite{Hatsuda:2019eoj}
\begin{eqnarray}
\omega^2 &=& V(r^*_{max})  -2 i U \,,
\end{eqnarray}
where

\begin{eqnarray}
\notag U &=&  N\sqrt{-V^{(2)}/2}+\frac{i}{64} \left( -\frac{1}{9}\frac{V^{(3)2}}{V^{(2)2}} (7+60N^2)+\frac{V^{(4)}}{V^{(2)}}(1+4 N^2) \right) \\
&& +\frac{N}{2^{3/2} 288} \Bigg( \frac{5}{24} \frac{V^{(3)4}}{(-V^{(2)})^{9/2}} (77+188N^2) 
 +\frac{3}{4} \frac{V^{(3)2} V^{(4)}}{(-V^{(2)})^{7/2}}(51+100N^2) \\ \notag  && +\frac{1}{8} \frac{V^{(4)2}}{(-V^{(2)})^{5/2}}(67+68 N^2)
+\frac{V^{(3)}V^{(5)}}{(-V^{(2)})^{5/2}}(19+28N^2)+\frac{V^{(6)}}{(-V^{(2)})^{3/2}} (5+ 4N^2)  \Bigg)\,,
\end{eqnarray}
and $N=n+1/2$, with $n=0,1,2,\dots$, is the overtone number.
The imaginary and real part of the QNFs can be written as
\begin{eqnarray}
\label{im} \omega_I^2 &=& - (Im(U)+V/2)+\sqrt{(Im(U)+V/2)^2+Re(U)^2} \,, \\
\omega_R^2 &=& -Re(U)^2 / \omega_I^2 \,,
\end{eqnarray}
respectively, where $Re(U)$ is the real part of $U$ and $Im(U)$ its imaginary part.
Now, defining $L^2= \ell (\ell+1)$, we find that for large values of $L$, the maximum of the potential is approximately at
\begin{equation}
r_{max} \approx r_{0}+\frac{1}{L^2}r_{2}\,,
\end{equation}
where $r_0$ is solution of equation
\begin{equation}
2 (r_0/r_h) - 3 + \sqrt{\tilde{B}+ \tilde{q}^2} + \tilde{q} \, {\rm arcsinh} \left( \frac{\tilde{q}}{\sqrt{\tilde{B}} (r_0/r_h )^3} \right) - \tilde{q} \, {\rm arcsinh} \left( \frac{\tilde{q}}{\sqrt{\tilde{B}}}  \right)  = 0 \,.
\end{equation}

Due to the impossibility of obtaining an analytical expression for $r_{0}$, from now on the position of the maximum will be expressed in terms of $r_{0}$ with

\small{
\begin{eqnarray}
\notag r_2 &=& - r_h \bigg(\tilde{r}_0^6 \tilde{B} \big( 4 \tilde{B} (\tilde{r}_0^6 +1)+ A C + 9 \tilde{r}_0 (-3+ A) - 3 \tilde{r}_0^3 \tilde{m}^2 (2 C -3 +A) - 3 (C-12+8A) \big)\\
\notag &&+q^2 \big( (3 \tilde{r}_0^6 +4)\tilde{B} +10 A C   
 - 3 \tilde{m}^2 \tilde{r}_0^3 (2 C -3 + A)  + 9 \tilde{r}_0 (3 C - 3 + A) - 6 (5 C -6 + 4A) \big) -\tilde{q}^4 \\
 && + \big( \tilde{r}_0^6  \tilde{q}  \tilde{B} (9 \tilde{r}_0 - 3 \tilde{r}_0^3 \tilde{m}^2 + C -24 + 8 A )  +  
 \tilde{q}^3 (9 \tilde{r}_0 - 3 \tilde{r}_0^3 \tilde{m}^2 +10 C -24 + 8 A )  \big) D \\
\notag &&+ 4 \tilde{q}^2 (\tilde{q}^2 + \tilde{r}_0^6 \tilde{B})D^2 \bigg) / \bigg( 9 \big( 2 \tilde{r}_0^6 \tilde{B} (3 \tilde{r}_0 - 6 +2 A) + \tilde{q}^2 (6 \tilde{r}_0 + 3 C -12 + 4 A)  \big) 
+36 \tilde{q} (\tilde{q}^2 + \tilde{B} \tilde{r}_0^6 ) D  \bigg) \,,
\end{eqnarray}
}
where
\begin{equation}
A = \sqrt{\tilde{q}^2+ \tilde{B} } \,, \, \, \, C =  \sqrt{\tilde{q}^2+ \tilde{B} \tilde{r}_0^6} \,, \, \, \, D = {\rm arcsinh}\left( \frac{\tilde{q}}{\sqrt{\tilde{B} \tilde{r}_0^3}} \right) - {\rm arcsinh}\left( \frac{\tilde{q}}{\sqrt{\tilde{B}}} \right) \,.
\end{equation}

So, the potential evaluated at  $r^*_{max}$, is given by
\begin{eqnarray}
r_h^2 V(r_{max}^*) &=& \frac{\tilde{r}_0-\sqrt{\tilde{q}^2 + \tilde{B} \tilde{r}_0^6}}{3 \tilde{r}_0^3} L^2 \\
\notag  &&+ \frac{2 \tilde{q}^2 + \tilde{r}_0 \left(2 \tilde{r}_0 + 3 \tilde{r}_0^3 \tilde{m}^2 + 2 \tilde{r}_0^5 \tilde{B} - 4 \sqrt{ \tilde{q}^2 + \tilde{B} \tilde{r}_0 ^6} - 3 \tilde{r}_0^2 \tilde{m}^2 \sqrt{ \tilde{q}^2 + \tilde{B} \tilde{r}_0^6} \right)}{9 \tilde{r}_0^4} + \mathcal{O}\left( L^{-2} \right) \,.
\end{eqnarray}

\begin{figure}[h]
\begin{center}
\includegraphics[width=0.4\textwidth]{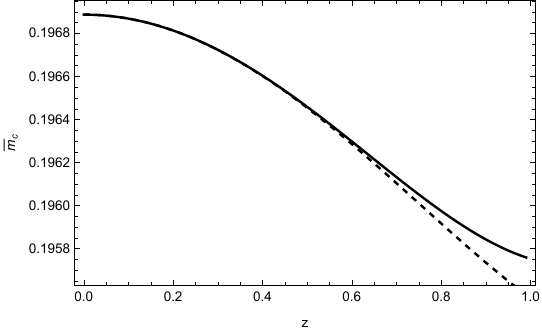}
\includegraphics[width=0.4\textwidth]{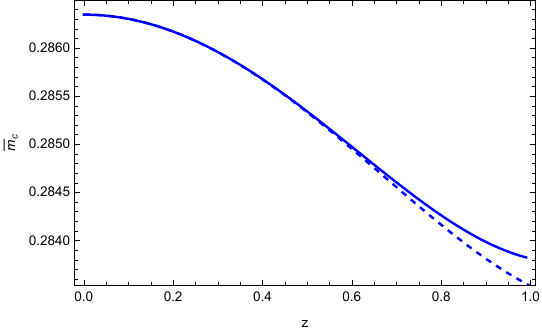}
\includegraphics[width=0.4\textwidth]{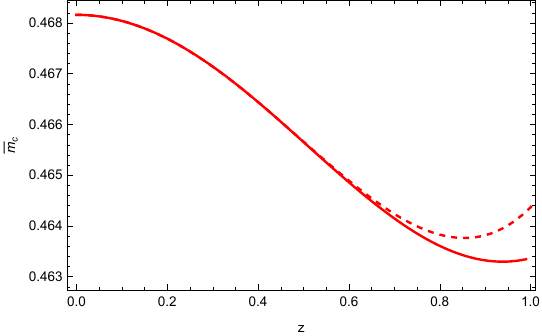}
\end{center}
\caption{The behaviour of the critical scalar field mass $\tilde{m}_{c}$ for the fundamental mode $n=0$ as a function of $z$. Black line for $\tilde{B}=0.001$, blue line for $\tilde{B}=0.01$, and red line for $\tilde{B}=0.1$. Here, the continuous line represents the analytical solution, and the dashed line represents the approximate solution given by Eq. (\ref{mcaprox}).}
\label{criticalmass}
\end{figure}

\begin{figure}[h]
\begin{center}
\includegraphics[width=0.4\textwidth]{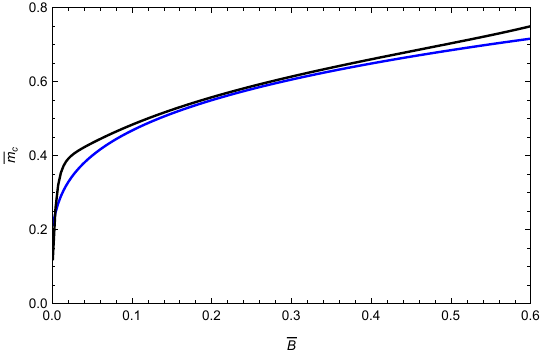}
\end{center}
\caption{The behaviour of the critical scalar field mass $\tilde{m}_c$ for the fundamental mode ($n=0$) as a function of $\tilde{B}$. Black line for $\tilde{q}= 0.6$ and blue line for $\tilde{q}=0.1$. Here, there is a good agreement between the analytical and the approximate solution for $\tilde{m}_{c}$ for small values of $\tilde{q}$ with $\tilde{B}$ large enough. The plot corresponds to the analytical solution.}
\label{cmB}
\end{figure}

The higher order derivatives  $V^{(i)}(r*_{max})$ with $i=2,...,6$  are not written here because are quite lenghty. Now, considering the following expansion of the QNFs for high values of $\ell$
\begin{equation}
\label{omegawkb}
\tilde{\omega}=\tilde{\omega}_{1m}L+\tilde{\omega}_{0}+\tilde{\omega}_{1}L^{-1}+\tilde{\omega}_{2}L^{-2} + \mathcal{O}(L^{-3})\,,
\end{equation}
we find
\begin{equation}
\tilde{\omega}_{1m} = \sqrt{ \frac{1}{3 \tilde{r}_0^2}- \frac{1}{3} \sqrt{\tilde{B} + \frac{\tilde{q}^2}{\tilde{r}_0^6}} } \,,
\end{equation}
\begin{equation}
\tilde{\omega}_0 = - \frac{i}{6 \sqrt{2} \tilde{\omega}_{1m}} \sqrt{\frac{2 ( 4 \tilde{q}^2 + \tilde{r}_0^2)}{\tilde{r}_0^6} - \frac{(3 \tilde{q}+ 7 \tilde{r}_0^2) \sqrt{\tilde{q}^2+ \tilde{B} \tilde{r}_0^6}}{\tilde{r}_0^7}+ \tilde{B} \left( 2 + \frac{ 3 \tilde{r}_0}{ \sqrt{\tilde{q}^2 + \tilde{B} \tilde{r}_0^6}} \right)} \,.
\end{equation}
The expressions for both $\tilde{\omega}_{1}$ and $\tilde{\omega}_2$ are in appendix \ref{FreqB}.
However, the above expressions predominate in the eikonal limit.
The critical scalar field mass, denoted as $\tilde{m}_c$, is defined by the mass value at which $\tilde{\omega}_2$ is nullified.  The complete analytical expression for $\tilde{m}_c$ is rather lengthy; however, for small values of the parameter $z=\tilde{q}/\sqrt{\tilde{B}}$, the critical mass $\tilde{m}_c$ can be approximated by expanding  it in terms of $z$ as follows
\begin{eqnarray}
\label{mcaprox}
\nonumber \tilde{m}_{c}&\approx& \frac{1}{18} K^{1/2}+\frac{ \sqrt{\tilde{B}} \left(5 \sqrt{\tilde{B}} \left(713 \tilde{B}-4689 \sqrt{\tilde{B}}+8883\right)-65863 \right) z^2}{450 \left(\sqrt{\tilde{B}}-3\right)^6  \sqrt{K}}+\\
\nonumber &&  \frac{\sqrt{\tilde{B}}}{270000 \left(\sqrt{\tilde{B}}-3\right)^{14} K^{3/2}}\Biggl(13864007728971 \sqrt{\tilde{B}} + 421198869100 + 5483223271446 \tilde{B}-
\\
\nonumber && 55996442329261 \tilde{B}^{3/2}+ 87627522635640 \tilde{B}^2-80661983356170 \tilde{B}^{5/2} - 24103194444450 \tilde{B}^{7/2}+\\
 &&
\nonumber 7851752187300 \tilde{B}^4 + 52007673850260 \tilde{B}^3+246140115750 \tilde{B}^5-1733501392225 \tilde{B}^{9/2}- \\
&& 20336564625 \tilde{B}^{11/2}+747865000 \tilde{B}^6  \Biggr) z^4 + \mathcal{O}(z^6) \,,
\end{eqnarray}
where
\begin{equation}
    K=\frac{401 \sqrt{\tilde{B}}}{2}+\frac{274}{5 \left(\sqrt{\tilde{B}}-3\right)^2}\,.
\end{equation}
It is worth noticing that the above expression for $z=0$ reduces to the critical mass of Schwarzschild-dS obtained in \cite{Aragon:2020tvq}.  This agrees with Eq. (\ref{appr}) which reduces to the Schwarzschil-dS metric for $z=0$. So, in order to visualize the range of values $z$, where the approximate solution is valid in comparison with the analytical solution, in Fig. \ref{criticalmass} we show the behaviour of  $\tilde{m}_{c}$ as a function of $z$ for both solutions, where the continuous line represents the analytical solution, and the dashed line represents the approximate solution given by Eq. (\ref{mcaprox}). Thus, for $\tilde{B}=0.001$,$\tilde{B}=0.01$ and  $\tilde{B}=0.1$ there is a very good agreement until $z\approx 0.6$. Also, we can observe that the parameter $\tilde{m}_c$ increases when the parameter $\tilde{B}$ increases. On the other hand, it is possible to observe that $\tilde{m}_{c}$ increases when the parameter $\tilde{B}$ increases, and when the parameter $\tilde{q}$ increases, see Fig. \ref{cmB}.


\begin{figure}[h]
\begin{center}
\includegraphics[width=0.4\textwidth]{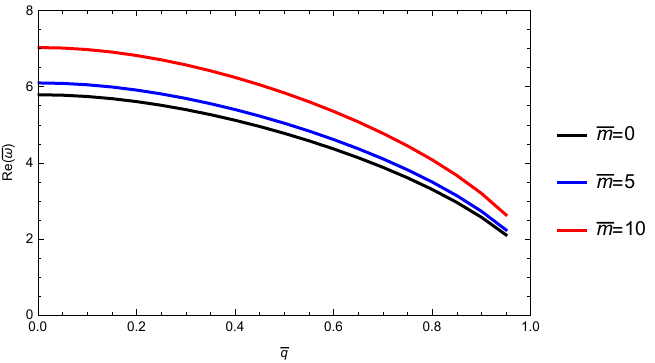}
\includegraphics[width=0.4\textwidth]{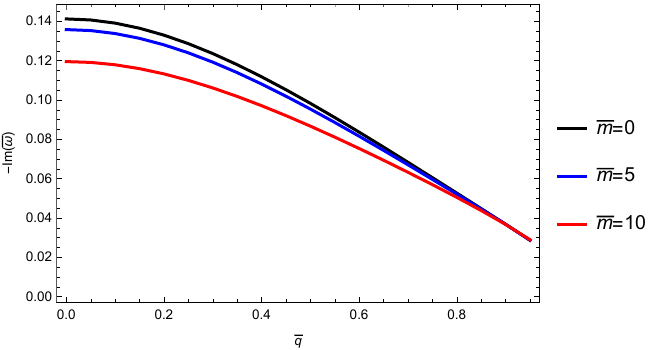}
\end{center}
\caption{The behaviour of $Re(\tilde{\omega})$ (left panel), and $-Im(\tilde{\omega})$ (right panel) for the fundamental mode ($n=0$) as a function of the  parameter $\tilde{q}$ with $\tilde{B}= 1/10$, and $\ell=20$.}
\label{ReIm}
\end{figure}

\begin{figure}[h]
\begin{center}
\includegraphics[width=0.4\textwidth]{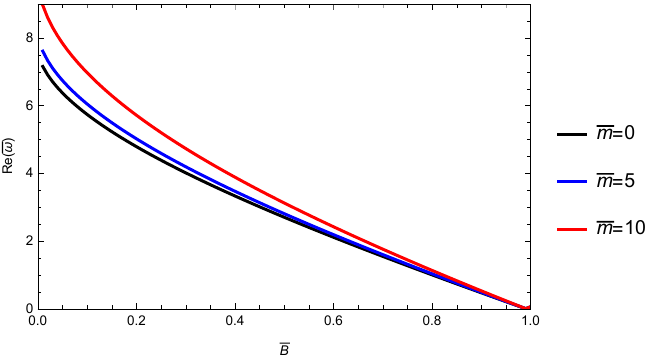}
\includegraphics[width=0.4\textwidth]{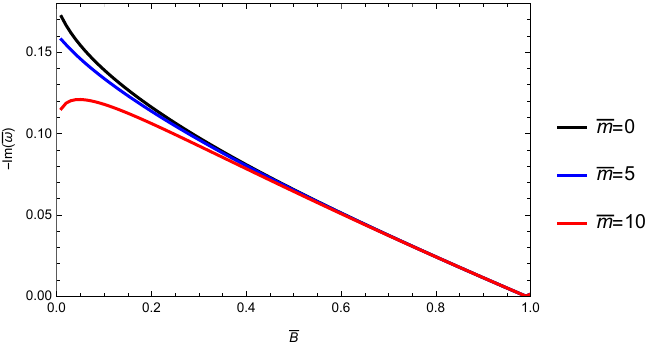}
\end{center}
\caption{The behaviour of $Re(\tilde{\omega})$ (left panel), and $-Im(\tilde{\omega})$ (right panel) for the fundamental mode ($n=0$) as a function of the  parameter $\tilde{B}$ with $\tilde{q}= 0.1$, and $\ell=20$.  }
\label{ReImB}
\end{figure}

In Fig.~\ref{ReIm}, we plot the behaviour of the real and imaginary parts of the QNFs as a function of $\tilde{q}$, we can observe that the oscillation frequency decreases when the parameter $\tilde{q}$ increases, and when the parameter $\tilde{m}$ decreases. Also, the absolute value of $Im(\tilde{\omega})$ decreases when the parameter $\tilde{q}$ increases. However, it tends to a constant value when $\tilde{q} \rightarrow 1$.
\newpage
Also, we show the behaviour of the real and imaginary parts of the QNFs as a function of $\tilde{B}$ in Fig. \ref{ReImB}. We can observe that the frequency of oscillation decreases when the parameter $\tilde{B}$ increases, and when the mass parameter $\tilde{m}$ decreases. Also, the absolute value of $Im(\tilde{\omega})$ decreases when the parameter $\tilde{B}$ increases, except for small values of $\tilde{B}$, and $\tilde{m}=10$. However, the real and the imaginary parts of the QNFs tends to a constant value when the black hole becomes extremal.

\begin{figure}[h]
\begin{center}
\includegraphics[width=0.32\textwidth]{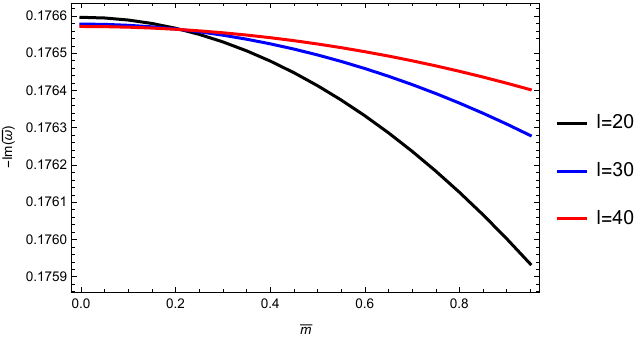}
\includegraphics[width=0.32\textwidth]{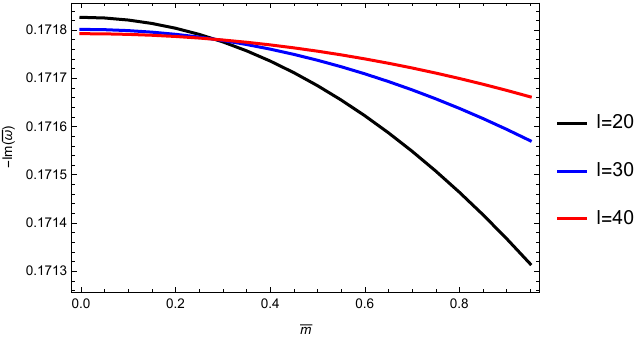}
\includegraphics[width=0.32\textwidth]{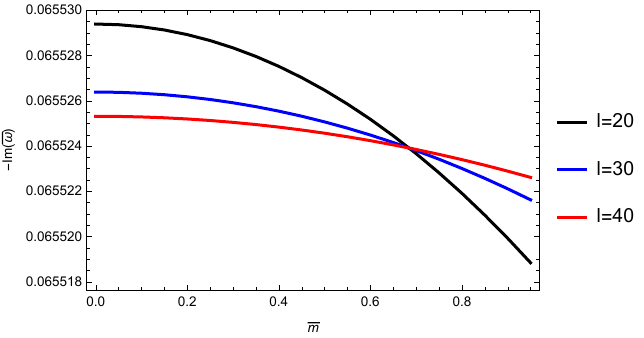}
\end{center}
\caption{The behaviour of $-Im(\tilde{\omega})$ for the fundamental mode ($n=0$) as a function of the scalar field mass $\tilde{m}$ for different values of the angular number $\ell=20,30,40$, with $\tilde{q}=0.1$, $\tilde{B}=0.001$ (left panel), $\tilde{B}=0.01$ (central  panel), and  $\tilde{B}=0.5$ (right panel). Here, the WKB method gives $\tilde{m}_{c}\approx0.212$, $\tilde{m}_{c}\approx0.284$, and $\tilde{m}_{c}\approx0.685$, respectively.}
\label{AB1}
\end{figure}

\begin{figure}[h]
\begin{center}
\includegraphics[width=0.32\textwidth]{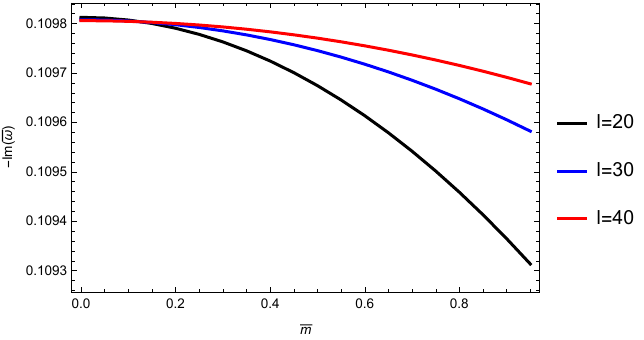}
\includegraphics[width=0.32\textwidth]{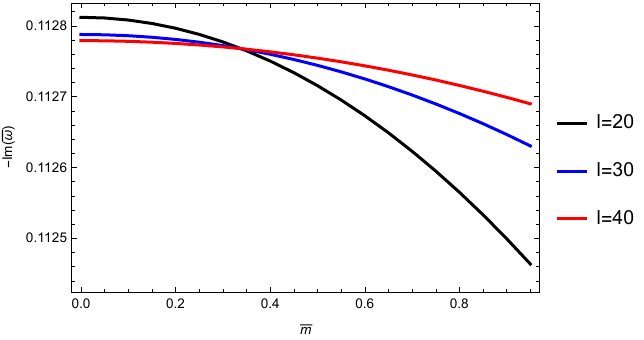}
\includegraphics[width=0.32\textwidth]{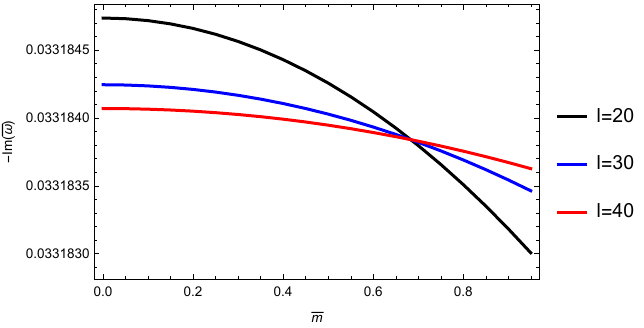}
\end{center}
\caption{The behaviour of $-Im(\tilde{\omega})$ for the fundamental mode ($n=0$) as a function of the scalar field mass $\tilde{m}$ for different values of the angular number $\ell=20,30,40$, with $\tilde{q}=0.5$, $\tilde{B}=0.001$ (left panel), $\tilde{B}=0.01$ (central panel), and  $\tilde{B}=0.5$ (right panel). Here, the WKB method gives $\tilde{m}_{c}\approx 0.117 $, $\tilde{m}_{c}\approx 0.337 $, and $\tilde{m}_{c}\approx 0.685 $, respectively.}
\label{AB2}
\end{figure}

Now, in order to show the anomalous behaviour, we plot in Fig. \ref{AB1}, and \ref{AB2}, the behaviour of $-Im(\tilde{\omega})$ as a function of $\tilde{m}$ by using the 3th order WKB  method for $n=0$, $\tilde{q}=0.1$, and $\tilde{q}=0.5$, respectively. We can observe an anomalous decay rate, i.e,  for $\tilde{m}<\tilde{m}_c$, the longest-lived modes are the ones with highest angular number $\ell$; whereas, for $\tilde{m}>\tilde{m}_c$, the longest-lived modes are the ones with smallest angular number. Also, it is possible to observe that the behaviour of the critical scalar field mass agrees with Fig. \ref{criticalmass}.

 As mentioned, the PS modes are determined by the behaviour of the effective potential near its maximum. They are represented by wave packets circulating near the photon sphere \cite{Cardoso:2008bp}, which penetrate the potential barrier and gradually leak out to the event horizon, where the tunneling probability according to the WKB formula depends on the potential width. Also the modes can escape to the spatial infinity or the cosmological horizon. In the eikonal limit, it can be shown that the imaginary part of the frequencies is inversely proportional to the width of the potential around its maximum \cite{Lagos:2020oek}: $Im(\omega)^2  \approx \frac{1}{4} \frac{(1- \epsilon)}{(\Delta r^{\ast})^2} + \mathcal{O}(L^{-2})$, where $\Delta r^{\ast}$ is the width of the potential when it has decayed from its maximum to $\epsilon V(r^{\ast}_{max})$.
We found that for low mass values, the width of the potential barrier increases with higher values of $\ell$, so making it more difficult for the waves to penetrate the potential barrier and leak to the event horizon, implicating that the longest-lived modes are those with the highest angular number $\ell$. Conversely, for high mass values, this behaviour is inverted. However, due to the presence of the higher order terms, specifically $\mathcal{O}(L^{-2})$, the behaviour inversion of $Im(\omega)$ does not align precisely with the mass value where the width of the potential changes its behaviour.
In Fig. \ref{pot} we show the behaviour of the effective potential near its maximum, therefore suffices to consider only up to second order in the Taylor expansion $\tilde{V} (\tilde{r}^{\ast}) \approx \tilde{V}(\tilde{r}^{\ast}_{max})+ \frac{1}{2!}\tilde{V}^{(2)}(\tilde{r}^{\ast})(\tilde{r}^{\ast}-\tilde{r}^{\ast}_{max})^2$, for $\tilde{m}=0.1$ (left panel) and $\tilde{m}=0.5$ (right panel) for $\ell =1$ and $\ell=10$. It can be observed that for $\tilde{m}<\tilde{m}_c$, the width of the potential increases with $\ell$, while for $\tilde{m}>\tilde{m}_c$, the width decreases with $\ell$.

\begin{figure}[h]
\begin{center}
\includegraphics[width=0.4\textwidth]{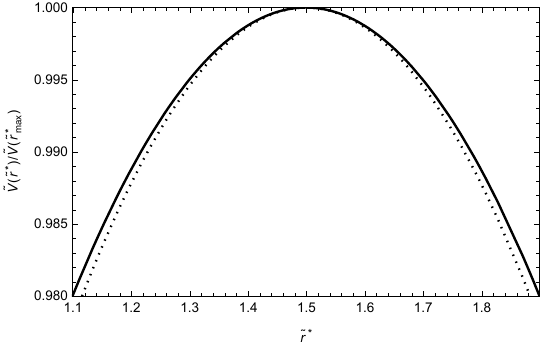}
\includegraphics[width=0.4\textwidth]{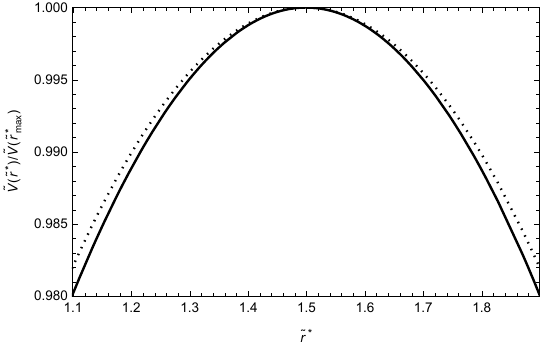}\\
\end{center}
\caption{The behaviour of the effective potential for $\tilde{q}=0.1$, 
$\tilde{B} = 0.001$ and $\tilde{m}=0.1$ (left panel), and $\tilde{m}=0.5$ (right panel) for $\ell =1$ (dotted line) and $\ell=10$ (solid line).}
\label{pot}
\end{figure}

\clearpage

\section{de Sitter modes}
\label{desitter}

The QNMs of a pure de Sitter spacetime are given by \cite{Du:2004jt}
\begin{equation}\label{deS}
\omega_{pure-dS} = - i \sqrt{\frac{\Lambda}{3}} \left( 2n_{dS} + \ell + 3/2 \pm \sqrt{\frac{9}{4}-3 \frac{m^2}{\Lambda}} \right)\,,
\end{equation}
where $\Lambda$ is the positive cosmological constant of pure de Sitter spacetime, $n_{dS}= 0, 1, 2, ...$ is the overtone number, and $m$ is the scalar field mass. It is worth to notice that for $m^2 \leq 3 \Lambda/4$ the QNFs of pure de Sitter spacetime are purely imaginary whereas for $m^2> 3\Lambda/4$ the QNFs acquire a real part. So, the QNMs for massive scalar perturbations in the background of black holes immersed in CDF with $\tilde{q}=0$ resemble those
of the pure de Sitter spacetime with  $\Lambda= \sqrt{\tilde{B}}/r_h^2$, and $m=\tilde{m}/r_h$. In Table \ref{QNMell} we show some QNFs for $\tilde{B}=0.005$ and for small values of the parameter $\tilde{q}$. So, we can observe that when $\tilde{q}$ decreases the modes resemble those of the pure de Sitter spacetime  as it is expected, $\tilde{\omega}_{pure-dS}=-0.1535260 i$ Eq. (\ref{deS}). Also, it is worth mentioning that the purely imaginary QNFs of asymptotically de Sitter black holes cannot be found by the standard WKB method; thereby we will use the pseudospectral Chebyshev method. Additionally, note that as the intensity parameter $\tilde{q}$ increases, the absolute value of the imaginary part decreases. Therefore, for the dS modes, the effect of the CDF is to result a propagation of massless scalar fields with a higher decay rate.


\begin {table}[ht]
\caption {QNFs ($\tilde{\omega}$) for massless scalar fields with $\ell=1$ in the background of black holes immersed in CDF, with $\tilde{B}= 0.005$ and $\tilde{q}= 0, 0.1, 0.2, 0.3$, and $0.4$. Here, the QNFs are obtained via the pseudospectral Chebyshev method using a number of Chebyshev polynomials in the range 95-100
with eight 
decimal places of accuracy for the QNF.}
\label {QNMell}\centering
\scriptsize
\begin {tabular} { | c |c |c |c |c |c |}
\hline
$n_{dS}$ &  $\tilde{q}=0$  &  $\tilde{q}=0.1$  & $\tilde{q}=0.2$ & $\tilde{q}=0.3$ & $\tilde{q}=0.4$ \\\hline
$0$ & $-0.15339081 i$ &
$-0.15338388 i$  & $-0.15336702 i$ & $-0.15334338 i $  & $ -0.15331391 i $
\\\hline
\end {tabular}
\end {table}


Now, in order to analyze the effect of the scalar field mass, we consider $\tilde{q}=0.1$ in Table \ref{dS1}, and $\tilde{m}=0, 0.1, 0.2, 0.3$. We can observe that for $\ell=0,1,2$, $n_{dS}=0$, and purely imaginary QNFs, the decay rate increases when the scalar field mass increases. Here, the zero mode
$\tilde{\omega}_{dS}=0$ with $n_{dS}=0$, and $\ell=0$, has been considered as a dS mode, as in Ref. \cite{Becar:2023jtd}. Note that the dS modes also can acquire a real part if the mass of the scalar field increases enough, which is similar to what happens to the modes of pure de Sitter spacetime. Also, to study the effects of the CDF, we present in Table \ref{dS1} the QNFs for the parameter $\tilde{q}=0$, representing Schwarzschild-dS black holes with $\Lambda=\sqrt{\tilde{B}}/r_h^2$. By comparing the results for $\tilde{q}=0$ and $\tilde{q}=0.1$, it is evident that both the real part of the QNFs and the absolute value of the imaginary part of the QNFs decrease as the parameter $\tilde{q}$ increases from $0$ to $0.1$. This indicates that the dS modes oscillate at lower frequencies and decay more rapidly for black holes immersed in a Chaplygin-like dark fluid.

\begin{table}[h]
\caption {de Sitter QNFs $\tilde{\omega}_{dS}$ for massive scalar fields in backgrounds of black holes immersed in CDF, with $\tilde{q}= 0, 0.1$. Here, the QNFs are obtained via the pseudospectral Chebyshev method using a number of Chebyshev polynomials in the range $95$-$100$, with eight  decimal places of accuracy for the QNFs.}
\label {dS1}\centering
\scriptsize
\begin {tabular} { | c | c | c | c | c |  }
\hline
\multicolumn{5}{ |c| }{$\tilde{q}=0.1$} \\ \hline
$\tilde{B}=0.001$ & $\tilde{m} = 0$ & $ \tilde{m} = 0.1$  & $ \tilde{m} = 0.2$
&  $ \tilde{m} = 0.3$
\\\hline
$\tilde{\omega}_{dS} ( \ell=0;n_{dS}=0)$ &
$0$ &
$-0.03506763 i$ &
$\pm 0.14431950-0.13656706 i$ &
$\pm 0.24971860-0.09389492 i $
 \\\hline
$\tilde{B}=0.0025$ & $\tilde{m} = 0$ & $ \tilde{m} = 0.1$  & $ \tilde{m} = 0.2$
&  $ \tilde{m} = 0.3$
\\\hline
$\tilde{\omega}_{dS} (l=0;n_{dS}=0)$ &
$0$ &
$-0.02567727 i$ &
$\pm 0.06911116-0.19389077 i$ &
$ \pm 0.24134731-0.23774836 i$
\\\hline
$\tilde{B}=0.1$ & $\tilde{m} = 0$ & $ \tilde{m} = 0.1$  & $ \tilde{m} = 0.2$
&  $ \tilde{m} = 0.3$
\\\hline
$\tilde{\omega}_{dS} (\ell=0;n_{dS}=0)$ &
$0$ &
$-0.00667282 i$ &
$-0.02879126 i$ &
$-0.07711202 i$
 \\\hline

\multicolumn{5}{ |c| }{$\tilde{q}=0$} \\ \hline
 $\tilde{B}=0.001$ & $\tilde{m} = 0$ & $ \tilde{m} = 0.1$  & $ \tilde{m} = 0.2$
&  $ \tilde{m} = 0.3$
\\\hline
$\tilde{\omega}_{dS} ( \ell=0;n_{dS}=0)$ &
$0$ &
$-0.03512118 i$ &
$\pm 0.14270000-0.13675999 i$ &
$\pm 0.25083520-0.09738286 i $
 \\\hline
$\tilde{B}=0.0025$ & $\tilde{m} = 0$ & $ \tilde{m} = 0.1$  & $ \tilde{m} = 0.2$
&  $ \tilde{m} = 0.3$
\\\hline
$\tilde{\omega}_{dS} (l=0;n_{dS}=0)$ &
$0$ &
$-0.02572495 i$ &
$\pm 0.06928080-0.19346048 i$ &
$ \pm 0.23701384-0.12065734 i$
\\\hline
$\tilde{B}=0.1$ & $\tilde{m} = 0$ & $ \tilde{m} = 0.1$  & $ \tilde{m} = 0.2$
&  $ \tilde{m} = 0.3$
\\\hline
$\tilde{\omega}_{dS} (\ell=0;n_{dS}=0)$ &
$0$ &
$-0.00669368 i$ &
$-0.02886091 i$ &
$-0.07713150 i$
 \\\hline
 
\end {tabular}
\end{table}\leavevmode\newline



\newpage
\section{Near extremal modes}
\label{nemodes}

When the Cauchy horizon $\tilde{r}_C$ approaches to the event horizon $\tilde{r}_h$ the family of modes that dominate is the near extremal family, which is characterized by purely imaginary frequencies.  In table \ref{NE1} we show the lowest near extremal QNFs for a massless scalar field, which decrease when the black hole approaches to the extremal limit
and in table \ref{NE3} the lowest QNFs for a massive scalar field.  Here, we observe that for $\ell =0$ the near extremal modes increase when the scaled mass $\tilde{m}$ increases. However, for $\ell=1$ it is not noted a difference in the frequencies up to four decimal places.

In the near extremal limit, for a massless scalar field this family tends to approach \cite{Cardoso:2017soq}
\begin{equation}
\label{new}
\omega_{NE} = - i (\ell + n + 1) \kappa_C = - i (\ell +n+1) \kappa_h  \,,
\end{equation}
where $\kappa_C = \frac{1}{2} |f'(r_C) |$ and $\kappa_h = \frac{1}{2} f'(r_h)$ are the surface gravity on the Cauchy horizon and event horizon, respectively.
\begin {table}[ht]
\caption {The lowest QNF $\tilde{\omega}_{NE}$ of the near extremal modes for massless scalar fields with $\ell=0$ in the background of black holes immersed in CDF, with $\tilde{B}= 0.01 $ and $\tilde{q}= 0.95,0.96, 0.97, 0.98, 0.99$. Here, the QNFs are obtained via the pseudospectral Chebyshev method using a number of Chebyshev polynomials in the range 95-100. The number of decimal places of accuracy for the QNFs is according to the precision achieved for the parameters considered.
The frequencies $\tilde{\omega}_{NE}$ have been obtained from Eq. (\ref{new}) at $r_C$, and scaling by the event horizon radius. 
}
\label{NE1}\centering
\scriptsize
\begin {tabular} { | c |c |c |c |c |c |}
\hline
{} & $\tilde{r}_C$   & $\Delta= \tilde{r}_h - \tilde{r}_C$  & $\tilde{\omega}$ & $\tilde{\omega}_{NE}$  \\\hline
$\tilde{q}= 0.95 $ & $0.909$  & $0.091$ & $-0.024427500 i$ &  $- 0.026617 i$
\\\hline
$\tilde{q}= 0.96$ & $0.930$ & $0.070$ & $-0.01860790 i$  &$ - 0.019272 i$
\\\hline
$\tilde{q}= 0.97$ & $0.950$  & $0.050$ & $-0.013028 i$ & $- 0.013165 i$
\\\hline
$\tilde{q}= 0.98$ & $0.969$  & $0.031$ & $-0.00767 i$ & $- 0.008105 i$
\\\hline
$\tilde{q}= 0.99$ & $0.990$ & $0.010$ & $-0.0025 i$ & $-0.0024 i$
\\\hline
\end {tabular}
\end {table}
\begin {table}[ht]
\caption {The lowest QNF $\tilde{\omega}_{NE}$ of the near extremal modes for massive scalar fields with $\ell=0, 1$ in the background of black holes immersed in CDF, with $\tilde{B}= 0.01 $, and $\tilde{q}= 0.95$. Here, the QNFs are obtained via the pseudospectral Chebyshev method using a number of Chebyshev polynomials in the range 95-100.
Here, the number of decimal places of accuracy for the QNFs is according to the precision achieved for the parameters considered. 
}
\label{NE3}\centering
\scriptsize
\begin {tabular} { | c| c |c |c |c |c |}
\hline
{} & $\tilde{m} =0$ & $\tilde{m}= 0.1$ &  $\tilde{m}=0.2$  & $\tilde{m}=0.3$ & $\tilde{m}=0.4$   \\\hline
$\tilde{\omega} (\ell=0)$ & $-0.024427500 i$ & $-0.02452894 i$ & $-0.0259 i$ & $-0.0259 i$  & $-0.0259 i$
\\\hline
$\tilde{\omega} (\ell=1)$ & $-0.02602 i$ & $-0.02602 i$  & $-0.0260 i$ & $-0.0260 i$ & $-0.02603 i$
\\\hline
\end {tabular}
\end {table}

It is worth mentioning that in this analysis, we have considered the near-extremal modes as the Cauchy and event horizons tend to coalesce, following the approach outlined in Ref. \cite{Cardoso:2017soq}. However, for the Schwarzschild-dS black hole, such coalescence does not occur because the Cauchy horizon is always negative and lacks physical significance. Consequently, we cannot analyze the effect of the Chaplygin-like dark fluid by setting the parameter $\tilde{q}=0$, as we have done for other families of modes. The near-extremal modes for Schwarzschild-dS black holes were analyzed in Ref. \cite{Cardoso:2003sw} by considering the convergence between the event and cosmological horizons, a consideration not addressed in this study.

\section{Dominance family modes}
\label{family}

As we mentioned the purely imaginary modes belong to the family of dS modes, and they continuously
approach those of pure de Sitter space in the limit that $\tilde{q}$ vanishes. However, the dS modes also can acquire a real part if the mass of the scalar field increases enough, which is similar to what happens to the modes of pure de Sitter spacetime. The other family corresponds to complex modes, for massless and massive scalar field, with a non null real part, namely PS modes.

\begin{figure}[h]
\begin{center}
\includegraphics[width=0.35\textwidth]{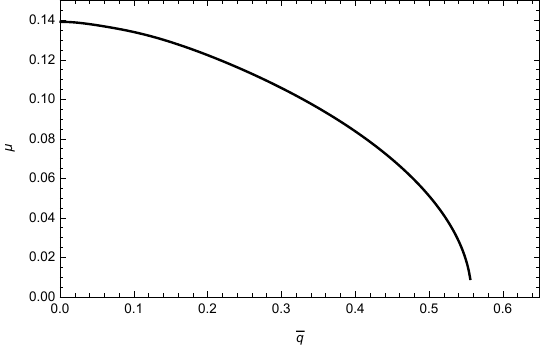}
\includegraphics[width=0.35\textwidth]{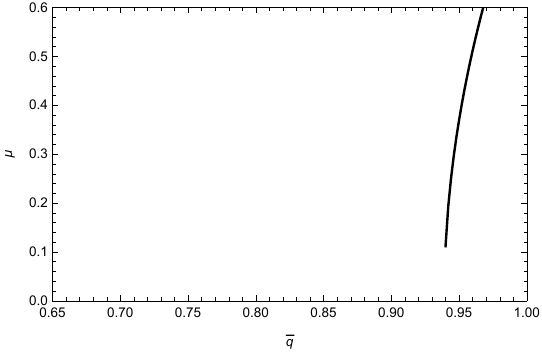}
\end{center}
\caption{The behaviour of $\mu$ as a function of $\tilde{q}$ for $\tilde{B}=0.001$, $\ell=1$. Here, $\mu \approx 0.131528 $ for $\tilde{q}=0.1$. The curve corresponds to $Im(\tilde{\omega}_{pure-dS})=Im(\tilde{\omega}_{PS})$ in the left panel, while in the right panel the curve corresponds to $Im(\tilde{\omega}_{PS}) = Im(\tilde{\omega}_{NE})$.}
\label{Dominance}
\end{figure}

Now, as in Ref. \cite{Becar:2023jtd}, in order to give an approximate value of $\tilde{m}=\mu$, where there is an interchange in the family dominance, we consider $Im(\tilde{\omega}_{dS})=Im(\tilde{\omega}_{PS})$ as a proxy for where the interchange in the family dominance occurs, where for $\tilde{\omega}_{PS}$ we consider the analytical expression via the WKB method, which yields the QNFs at third order beyond the eikonal limit, and for $\tilde{\omega}_{dS}$ we consider the analytical expression given by Eq. (\ref{deS}) for the pure de Sitter spacetime $\omega_{pure-dS}$ with $\Lambda = \sqrt{\tilde{B}}/r_h^2$, which yields well-approximated QNFs for the dS family for high values of $\ell$ or small values $\tilde{q}$. It is important because allows discern if the dominant family is able to suffers the anomalous behaviour of the decay rate. Despite for the
background considered is not possible to give an analytical expression for $\mu$, we consider the following parameter $\tilde{B}=0.001$, $\ell=1$, and $n_{PS}=n_{dS}=0$.
So, the equality of $Im(\omega_{pure-dS})$ with  $Im(\tilde{\omega}_{PS})$ is shown in Fig. \ref{Dominance} (left panel), where the line separates regions in the parameters space where a family of QNFs dominates according to the analytical approximation. Note that $\mu$ decreases when the intensity of CDF increases for the parameters considered. In the region down the line always the de Sitter modes dominate, while that in the region above it the PS modes dominate, which is validated in Table \ref{Dq} for $\tilde{q} = 0,\, 0.1$, and $\ell=1$, via the pseudospectral Chebyshev method, where the purely imaginary QNFs belong to the dS family, and the complex QNFs belong to the PS family. Also, we show that when the PS modes dominate for massless scalar field these are still dominant for massive scalar field, for  $\ell=2$. It is worth mentioning that for $\ell>2$, the dominant modes are the PS modes, for the parameters considered.  Also, in Fig. 10 (right panel) we consider $Im(\tilde{\omega}_{PS}) = Im(\tilde{\omega}_{NE})$ as a proxy for where there is an interchange in the dominance of the PS modes and the near extremal modes, where for $\tilde{\omega}_{PS}$ we consider the analytical expression of the QNFs via the WKB method at third order beyond the eikonal limit, and for $\tilde{\omega}_{NE}$ we consider the analytical expression given by Eq. (\ref{new}). To the left of the curve the PS modes dominate while to the right the near extremal modes dominate.

\begin {table}[ht]
\caption {The fundamental QNF $\tilde{\omega}$ for massive scalar fields with $\ell=1,2$ in the background of black holes immersed in CDF, with $\tilde{B}= 0.001$, $\tilde{q}= 0, 0.1$. Here, the QNFs are obtained via the pseudospectral Chebyshev method using a number of Chebyshev polynomials in the range 95-100
with six decimal places of accuracy for the QNF.}
\label {Dq}\centering
\resizebox{1.0\textwidth}{!}{
\scriptsize
\begin {tabular} { | c |c |c |c |c |}
\hline
\multicolumn{5}{ |c| }{$\tilde{q}=0.1$}\\\hline
$\ell=1$ & $\tilde{m}=0.12$ &  $\tilde{m}=0.13$  & $\tilde{m}=0.14$ & $\tilde{m}=0.15$ \\\hline
$\tilde{\omega}$ & -0.159013 i  & -0.173083 i   & $\pm$ 0.558383-0.179282 i  & $\pm$ 0.559028 - 0.178930 i\\\hline
$\ell=2$ & $\tilde{m}=0$ &  $\tilde{m}=0.2$  & $\tilde{m}=0.4$ & $\tilde{m}=0.6$ \\\hline
$\tilde{\omega}$ & $\pm$ 0.917487 - 0.178566  i  & $\pm$ 0.923598 - 0.176675 i  & $\pm$ 0.942030 - 0.170960 i  & $\pm$ 0.973101 - 0.161280  i
\\\hline
\multicolumn{5}{ |c| }{$\tilde{q}=0$}\\\hline
$\ell=1$ & $\tilde{m}=0.12$ &  $\tilde{m}=0.13$  & $\tilde{m}=0.14$ & $\tilde{m}=0.15$ \\\hline
$\tilde{\omega}$ & -0.159054 i  & -0.173117 i   & $\pm$ 0.571549 - 0.190210 i  & $\pm$ 0.572191 - 0.189846 i \\\hline
$\ell=2$ & $\tilde{m}=0$ &  $\tilde{m}=0.2$  & $\tilde{m}=0.4$ & $\tilde{m}=0.6$ \\\hline
$\tilde{\omega}$ & $\pm$ 0.940207 - 0.189532 i & $\pm$ 0.946329 - 0.187570 i  & $\pm$ 0.964781 - 0.181651 i  & $\pm$ 0.995843 - 0.171662 i
\\\hline
\end {tabular}
}
\end {table}


On the other hand, we can observe that the dominance depends on the parameter $\tilde{B}$, for small values of $\tilde{B}$, the dominant family is the dS; however, when the parameter increases the dominant family is the PS, see Table \ref{DB}.

\begin {table}[ht]
\caption {The fundamental QNF $\tilde{\omega}$ for massless scalar fields with $\ell=1,2$ in the background of black holes immersed in CDF, with $\tilde{B}= 0.001, 0.0025, 0.005, 0.01, 0.1 $, $\tilde{q}= 0, 0.1$. Here, the QNFs are obtained via the pseudospectral Chebyshev method using a number of Chebyshev polynomials in the range 95-100
with six decimal places of accuracy for the QNF.
}
\label {DB}\centering
\resizebox{1.0\textwidth}{!}{
\scriptsize
\begin {tabular} { | c |c |c |c |c |c |}
\hline
\multicolumn{6}{ |c| }{$\tilde{q}=0.1$}\\\hline
{} & $\tilde{B}=0.001$ &  $\tilde{B}=0.0025$  & $\tilde{B}=0.005$ & $\tilde{B}=0.01$  & $\tilde{B}=0.1$  \\\hline
$\tilde{\omega} (\ell=1)$ & -0.102622 i & -0.129011 i & -0.153384 i & $\pm$ 0.519752 - 0.180111 i   & $\pm$ 0.395902 - 0.149061 i
\\\hline
$\tilde{\omega} (\ell=2)$ & $\pm$ 0.917487 - 0.178566 i & 0.906352 - 0.178576 i & 0.891899 - 0.177494 i & $\pm $ 0.869609 - 0.174847 i  & $\pm$ 0.685725 - 0.142303 i
\\\hline
\multicolumn{6}{ |c| }{$\tilde{q}=0$}\\\hline
{} & $\tilde{B}=0.001$ &  $\tilde{B}=0.0025$  & $\tilde{B}=0.005$ & $\tilde{B}=0.01$  & $\tilde{B}=0.1$  \\\hline
$\tilde{\omega} (\ell=1)$ & -0.102626 i & -0.129016 i & -0.153391 i & $\pm$ -0.182358 i   & $\pm$ 0.399052 - 0.151465 i
\\\hline
$\tilde{\omega} (\ell=2)$ & $\pm$ 0.940207 - 0.189532 i & $\pm$ 0.924415 - 0.187105 i & $\pm$ 0.906571 - 0.184275 i & $\pm $ 0.881253 - 0.180110 i  & $\pm$ 0.691200 - 0.144532 i
\\\hline
\end {tabular}
}
\end {table}




\clearpage
\section{Conclusions}
\label{conclusion}

In this work, we studied the propagation of massive scalar fields  in the background of BHs immersed in Chaplygin like dark fluid through the QNFs by using the pseudospectral Chebyshev and the WKB method. The QNMs are characterized by two families of modes. One of them is the PS modes, which are complex, and the other one corresponds to the dS modes, which for small values of the mass are purely imaginary, and can adquire a real part when the scalar field mass increases. The propagation of scalar fields in this background result stable for both families.

Concerning to the photon sphere modes and massless scalar fields propagation we have shown that the longest-lived modes are the ones with highest angular number $\ell$. The frequency of oscillation increases when the angular number $\ell$ increases. Also, when the energy density  of the CDF increases i.e., when the intensity parameter of the CDF $\tilde{q}$ or the $\tilde{B}$ parameter increases, the frequency of oscillation, and the absolute value of the imaginary part of the QNFs  decreases. Therefore, the effect of the CDF is to decreases the frequency of oscillation and to increases the time of the decay of the modes in comparison with the  Schwarzschild dS spacetime.

However, when the scalar field acquires mass we showed the existence of anomalous decay rate of QNMs. In other words, the absolute values of the imaginary part of the QNFs decrease as the angular harmonic numbers increase if the mass of the scalar field is smaller than a critical mass. On the contrary  they grow when the angular harmonic numbers increase, if the mass of the scalar field is larger than the critical mass, a behaviour observed in other dS spacetimes. Here, the values of the critical mass for a fix value $\tilde{q}$ increases when the paramater $\tilde{B}$ increases (Fig. \ref{criticalmass}, and \ref{cmB}).

Moreover, the frequency of oscillation decreases as the parameter $\tilde{q}$ increases, and as the parameter $\tilde{m}$ decreases, and the absolute values of $Im(\tilde{\omega})$ decreases when the parameter $\tilde{q}$ increases. However it tends to a constant value as $\tilde{q} \rightarrow 1$ (Fig. \ref{ReIm}). Additionally, the frequency of oscillation decreases as the parameter $\tilde{B}$ increases, and as the parameter $\tilde{m}$ decreases. Furthermore, the absolute value of $Im(\tilde{\omega})$ decreases as $\tilde{B}$ increases, except for small values of $\tilde{B}$, and {\bf{$\tilde{m}=10$}}. Nonetheless, the real and the imaginary parts of the QNFs tends to a constant value when the black hole becomes extremal $\tilde{B} \rightarrow 1$ (Fig. \ref{ReImB}).

Regarding to the dS modes, we have shown when $\tilde{q}$ decreases the modes resemble those of the pure de Sitter spacetime. Additionally, when the parameter $\tilde{q}$ increases the absolute value of the imaginary part decreases. Therefore, the effect of the CDF is to increases the time of the decay of the modes in comparison with the Schwarzschild dS spacetime. Furthermore, when the scalar field acquires mass and it increases, the QNFs acquire a real part, similar to the pure de Sitter background.

Concerning to the dominance family modes we have shown that when the dS modes dominate for massless scalar field, which occurs for small values of the angular momentum $\ell=0,1$, and small $\tilde{B}$, there is a critical value beyond that the PS family dominate for $\tilde{B}$, or $\tilde{m}$. Additionally, we have shown that the PS modes  become dominant when the parameter $\ell$ increases. On the other hand, when the PS modes dominate for massless scalar field these are still dominant for massive scalar field. Also, we have shown that when the black hole approaches to the extremal limit the near extremal family becomes dominant, for $\tilde{B} = 0.001$ this happens at $\tilde{q} \sim 1$.
\newpage

\appendix
\section{Comparison of  Pseudospectral Chebyshev and WKB methods}
\label{Accuracy}

In Table \ref{TableI} we show the fundamental QNFs and the first overtone calculated using the pseudospectral Chebyshev method and the WKB approximation, and we see a good agreement for high $\ell$. We show the relative error of the real and imaginary parts of the values obtained with the WKB method with respect to the values obtained with the pseudospectral Chebyshev method, which is defined by
\begin{eqnarray}
\label{E}
\epsilon_{Re(\omega)} &=& \frac{\mid Re(\omega) - Re(\omega)_{\text{WKB}} \mid}{\mid Re(\omega \mid } \cdot 100\%\,, \\
\epsilon_{Im(\omega)} &=&\frac{\mid Im(\omega) - Im(\omega)_{\text{WKB}}\mid}{\mid Im(\omega)\mid } \cdot 100\% \,,
\end{eqnarray}
where $\omega$ denotes the result with the pseudospectral Chebyshev method, and $\omega_{WKB}$ corresponds to the result obtained with the third order WKB method.  We can observe that the error does not exceed 0.277 $\%$ in the imaginary part, and  0.103 $\%$ in the real part, for low values of $\ell=1,2,3,5$. However, for high values of $\ell$ ($\ell = 10, 15$), the error does not exceed  1.608$\cdot$10$^{-4}$ $\%$  in the imaginary part, and 2.170$\cdot$10$^{-5}$ $\%$ in the real part. Also, as it was observed, the frequencies all have a negative imaginary part, which means that the propagation of massive scalar fields is stable in this background.

\begin{table}[h]
\centering
\caption{The fundamental ($n=0$) QNFs ($\tilde{\omega}$) for several values of the angular momentum $\ell$ of the scalar field with $\tilde{m}=0,0.1$, for black holes immersed in CDF with $\tilde{q}=0.25$, and $\tilde{B}=0.1$, using the pseudospectral Chebyshev method and the third order WKB approximation. Here, the QNFs via the pseudospectral Chebyshev method have been obtained using a number of Chebyshev polynomials in the range $95$-$100$, with night or eight decimal places of accuracy.}
\scriptsize
\begin{tabular}{|c|c|c|c|c|}  \hline
\multicolumn{5}{|c|}{$\tilde{m}=0$}  \\  \hline
$\ell$ &  Pseudospectral Chebyshev method & WKB & $\epsilon_{Re(\tilde{\omega})}(\%)$ & $\epsilon_{Im(\tilde{\omega})}(\%)$ \\
\hline
  1  & $\pm$ 0.380134370 - 0.137398949 i & $\pm$ 0.379744439 - 0.137018184 i & 0.103 & 0.277 \\
  2  & $\pm$ 0.658353017 - 0.131451768 i & $\pm$ 0.658305481 - 0.131398216 i & 0.007 & 0.041 \\
  3   & $\pm$ 0.931201650 - 0.130008109 i & $\pm$ 0.931185647 - 0.129993224 i & 0.002 & 0.011 \\
    5  & $\pm$ 1.472528714 - 0.129152845 i & $\pm$ 1.472524573 - 0.129150229 i & 2.812$\cdot$10$^{-4}$ & 2.026$\cdot$10$^{-3}$ \\

   10  & $\pm$ 2.819847628 - 0.128741711 i & $\pm$ 2.819847016 - 0.128741504 i & 2.170$\cdot$10$^{-5}$ & 1.608$\cdot$10$^{-4}$ \\
  15  & $\pm$ 4.165243420 - 0.128658526 i & $\pm$ 4.165243228 - 0.128658482 i & 4.610$\cdot$10$^{-6}$ & 4.420$\cdot$10$^{-5}$ \\
\hline
\multicolumn{5}{|c|}{$\tilde{m}=0.1$} \\ \hline
  1  & $\pm$ 0.381767894 - 0.136924586 i  & $\pm$ 0.381475773 - 0.136626033 i & 0.077 & 0.218 \\
  2  & $\pm$ 0.65933996  -  0.13131186 i & $\pm$ 0.65930507 - 0.13126750 i & 0.005 & 0.034 \\
  3   & $\pm$ 0.931904467 - 0.129940465 i & $\pm$ 0.931892462 - 0.129927866 i & 0.001 & 0.010 \\
    5  & $\pm$ 1.472974799 - 0.129126333 i  & $\pm$ 1.472971602 - 0.129124086 i & 2.170$\cdot$10$^{-4}$ & 0.002 \\

   10  & $\pm$ 2.820080951 - 0.128734554 i & $\pm$ 2.820080469 - 0.128734374 i  & 1.709$\cdot$10$^{-5}$ & 1.398$\cdot$10$^{-4}$ \\
  15  & $\pm$ 4.165401428 - 0.128655253 i & $\pm$ 4.165401276 - 0.128655214 i & 3.649$\cdot$10$^{-6}$ & 3.031$\cdot$10$^{-5}$ \\
\hline

\end{tabular}
\label{TableI}
\end{table}

\clearpage

\section{Expressions for $\tilde{\omega}_{1}$ and $\tilde{\omega}_2$}
\label{FreqB}

In this appendix we show the analytical expressions for $\tilde{\omega}_{1}$ and $\tilde{\omega}_2$ of Eq. (\ref{omegawkb}).

\footnotesize

\begin{eqnarray}
\notag \tilde{\omega}_1 &=&  \Bigg( -486 \left(\tilde{B} \left(-\frac{3 \tilde{r}_0}{C}-2\right)+\frac{\left(3 \tilde{q}^2+7 \tilde{r}_0^2\right) C}{\tilde{r}_0^7}-\frac{2 \left(4 \tilde{q}^2+\tilde{r}_0^2\right)}{\tilde{r}_0^6}\right)^3  -\frac{11 \left(\tilde{r}_0-C\right)^3}{\tilde{r}_0^{21} C^8} \\
\notag && \Bigg( \tilde{q}^6 \left(27 C+58 \left(A-3\right)+50 \tilde{r}_0\right)-12 \tilde{B}^2 \tilde{r}_0^{14} \left(-2 C+\tilde{B} \tilde{r}_0^5+\tilde{r}_0\right) +
  58 \tilde{q}^3  D \left(2 \tilde{B} \tilde{r}_0^6+\tilde{q}^2\right)\\ 
\notag && \left(\tilde{r}_0 \left(-2 C+\tilde{B} \tilde{r}_0^5+\tilde{r}_0\right)+\tilde{q}^2\right)  + 
 2 \tilde{B} \tilde{q}^2 \tilde{r}_0^7 \Bigg(\tilde{B} \tilde{r}_0^5 \left(27 C+58 \left(A-3\right)\right)+\tilde{r}_0 \left(58 \left(A-3\right)-181 C\right)- \\
\notag   && 116 C\left(A-3\right)+44 \tilde{B} \tilde{r}_0^6+104 \tilde{r}_0^2\Bigg) +\tilde{q}^4 \tilde{r}_0 \Bigg(3 \tilde{B} \tilde{r}_0^5 \Bigg(27 C+ 
 58 \left(A-3\right)\Bigg)+\tilde{r}_0 \left(58 \left(A-3\right)-181 C\right)\\
\notag &&-116 C\Bigg(A- 3 \Bigg)+150 \tilde{B} \tilde{r}_0^6+104 \tilde{r}_0^2\Bigg)\Bigg)^2 + 81 \frac{1}{\tilde{r}_0^{14} C^6} \left(\tilde{r}_0-C\right) \left(C-\tilde{r}_0\right) \\
\notag &&  \left(\tilde{B} \left(-\frac{3 \tilde{r}_0}{C}-2\right)+\frac{\left(3 \tilde{q}^2+7 \tilde{r}_0^2\right) C}{\tilde{r}_0^7}-\frac{2 \left(4 \tilde{q}^2+\tilde{r}_0^2\right)}{\tilde{r}_0^6}\right) \Bigg( -48 \tilde{B}^3 \bigg(\tilde{B} \tilde{r}_0^5+\tilde{r}_0- 
 2 C\bigg) \tilde{r}_0^{21}\\ 
\notag && +12 \tilde{B}^2 \tilde{q}^2 \Bigg(5 \tilde{B}^2 \tilde{r}_0^{10}-247 \tilde{B} \tilde{r}_0^6+\tilde{B} \Bigg(-111 A+71 C +333\Bigg) \tilde{r}_0^5 +
 37 \tilde{B} C\left(A-3\right) \tilde{r}_0^4-86 \tilde{r}_0^2\\
\notag &&+\left(-37 A+253 C +111\right) \tilde{r}_0+
 111 C\left(A-3\right)\Bigg) \tilde{r}_0^{14}\\
\notag && -3 \tilde{B} \tilde{q}^4 \Bigg(\tilde{B}^2 \tilde{r}_0^{10}+2151 \tilde{B} \tilde{r}_0^6+\tilde{B} \bigg(888 \left(A-3\right) 
 -763 C\bigg) \tilde{r}_0^5-296 \tilde{B} C\left(A-3\right) \tilde{r}_0^4+344 \tilde{r}_0^2\\
\notag && +\Bigg(148 \left(A-3\right)  
 -1085 C\Bigg) \tilde{r}_0+444 C\left(3-A\right)\Bigg) \tilde{r}_0^8+\tilde{q}^6 \Bigg(-288 \tilde{B}^2 \tilde{r}_0^{10}-4851 \tilde{B} \tilde{r}_0^6 \\
\notag && +\tilde{B} \left(2075 C-1776 \left(A-3\right)\right) \tilde{r}_0^5+592 \tilde{B} C\left(A-3\right) \tilde{r}_0^4 
 -344 \tilde{r}_0^2+2 \left(-74 A+561 C+222\right) \tilde{r}_0\\
\notag && +444 C\left(A-3\right)\Bigg) \tilde{r}_0^2 
 -102 \tilde{q}^{10}+\tilde{q}^8 \Bigg(-327 \tilde{B} \tilde{r}_0^6-1314 \tilde{r}_0^2+\left(-444 A+638 C+1332\right) \tilde{r}_0\\
\notag && +148 C \Bigg( 
 A-3 \Bigg)\Bigg)+148  \tilde{q}^3 D \Bigg( 
\left(C-3 \tilde{r}_0\right) \tilde{q}^2+\left(-3 \tilde{B} \tilde{r}_0^5+\tilde{B} C \tilde{r}_0^4-\tilde{r}_0+3 C\right) \tilde{r}_0^2\Bigg)\\ 
\notag && \left(3 \tilde{B}^2 \tilde{r}_0^{12}+3 \tilde{B} \tilde{q}^2 \tilde{r}_0^6+\tilde{q}^4\right)  \Bigg) + \\
\notag && \Bigg( 1296 \left(\tilde{r}_0-C\right) \left(\tilde{B} \left(-\frac{3 \tilde{r}_0}{C}-2\right)+\frac{\left(3 \tilde{q}^2+7 \tilde{r}_0^2\right) C}{\tilde{r}_0^7}-\frac{2 \left(4 \tilde{q}^2+\tilde{r}_0^2\right)}{\tilde{r}_0^6}\right){}^2 \Bigg(\tilde{r}_0 \Bigg(\tilde{r}_0  \\
\notag && \left(3 m^2 \tilde{r}_0 \left(\tilde{r}_0-C\right)+2 \tilde{B} \tilde{r}_0^4+2\right)-4 C\Bigg)+2 \tilde{q}^2\Bigg) \Bigg) \bigg/ \tilde{r}_0^7 \Bigg) \bigg/ \Bigg( 7776 \sqrt{3} \left(\frac{\tilde{r}_0-C}{\tilde{r}_0^3}\right)^{3/2} \\
\notag && \left(\tilde{B} \left(-\frac{3 \tilde{r}_0}{C}-2\right)+\frac{\left(3 \tilde{q}^2+7 \tilde{r}_0^2\right) C}{\tilde{r}_0^7}-\frac{2 \left(4 \tilde{q}^2+\tilde{r}_0^2\right)}{\tilde{r}_0^6}\right)^2 \Bigg)
\end{eqnarray}

\clearpage

\begin{eqnarray}
\notag \tilde{\omega}_2 &=& \frac{6561 i \sqrt{\frac{3}{2}} \tilde{r}_0^{15/2}}{256 \sqrt{-\Phi } \Phi ^4 \left(\tilde{r}_0-C\right)^{5/2}} \Bigg(    -\frac{7 \Xi ^2 \Phi ^2 \left(\tilde{r}_0-C\right)^4}{531441 \tilde{r}_0^{28} C^{12}}+\frac{38 \Xi  \Upsilon ^2 \Phi  \left(\tilde{r}_0-C\right)^2}{4782969 \tilde{r}_0^{35} C^{14}}  
 \left(C-\tilde{r}_0\right)^3 \\
 \notag && -\frac{4 \Xi  \Phi ^4 \left(\tilde{r}_0-C\right) \left(C-\tilde{r}_0\right)}{59049 \tilde{r}_0^{14} C^6} -\frac{155 \Upsilon ^4 \left(\tilde{r}_0-C\right)^6}{387420489 \tilde{r}_0^{42} C^{16}}+  \frac{44 \Upsilon ^2 \Phi ^3 \left(\tilde{r}_0-C\right)^3}{4782969 \tilde{r}_0^{21} C^8}+\frac{8 \Phi ^6}{19683}\\
\notag && + 52 \Upsilon  \Phi ^2 \frac{\left(\tilde{r}_0-C\right){}^4}{4782969 \tilde{r}_0^{28} C^{12}} \Bigg( -360 \tilde{B}^4 \Bigg(3 \tilde{B} \tilde{r}_0^5-  \tilde{B} C \tilde{r}_0^4+\tilde{r}_0-3 C\Bigg) \tilde{r}_0^{28}\\
\notag && -24 \tilde{B}^3 \tilde{q}^2 \Bigg(84 \tilde{B}^2 \tilde{r}_0^{10}+\tilde{B}^2 \left(7 A+5 C-21\right) \tilde{r}_0^9  +559 \tilde{B} \tilde{r}_0^6+7 \tilde{B} \left(6 A-53 C-18\right) \tilde{r}_0^5-28 \tilde{B} C\left(A-3\right) \tilde{r}_0^4 \\
\notag && +74 \tilde{r}_0^2+\left(7 A-321 C-21\right) \tilde{r}_0+84 C-28 AC\Bigg) \tilde{r}_0^{21}  +6 \tilde{B}^2 \tilde{q}^4 \Bigg(-2720 \tilde{B}^2 \tilde{r}_0^{10}+5 \tilde{B}^2 \left(60 \left(A-3\right)+103 C\right) \tilde{r}_0^9\\
\notag && -2704 \tilde{B} \tilde{r}_0^6+2 \tilde{B} \Bigg(984 \bigg(A  -3\bigg)+1943 C\Bigg) \tilde{r}_0^5-1312 \tilde{B} C\left(A-3\right) \tilde{r}_0^4+352 \tilde{r}_0^2+ \\
\notag && \Bigg( 356 \left(A-3\right)-673 C\Bigg) \tilde{r}_0+1424 C\left(3-A\right)\Bigg) \tilde{r}_0^{15}+    \tilde{B} \tilde{q}^6 \Bigg(-38783 \tilde{B}^2 \tilde{r}_0^{10}+4 \tilde{B}^2 \left(1382 \left(A-3\right)+2499 C\right) \tilde{r}_0^9\\
\notag && -8589 \tilde{B} \tilde{r}_0^6+  \tilde{B} \left(21360 \left(A-3\right)+29201 C\right) \tilde{r}_0^5-14240 \tilde{B} C\left(A-3\right) \tilde{r}_0^4  +1408 \tilde{r}_0^2+\left(1424 \left(A-3\right)-2401 C\right) \tilde{r}_0\\
\notag && +5696 C\left(3-A\right)  \Bigg) \tilde{r}_0^9+\tilde{q}^8 \Bigg(-40977 \tilde{B}^2 \tilde{r}_0^{10}+3 \tilde{B}^2 \left(1780 \left(A-3\right)+3821 C\right) \tilde{r}_0^9-6798 \tilde{B} \tilde{r}_0^6+ \\
\notag && 8 \tilde{B} \left(1335 \left(A-3\right)+2459 C\right) \tilde{r}_0^5-7120 \tilde{B} C\left(A-3\right) \tilde{r}_0^4+    352 \tilde{r}_0^2+4 \left(89 A-103 C-267\right) \tilde{r}_0+1424 C\left(3-A\right)\Bigg) \tilde{r}_0^3 \\
\notag && +\tilde{q}^{10} \Bigg(-20805 \tilde{B} \tilde{r}_0^6+6 \tilde{B} \left(356 \left(A-3\right)+983 C\right) \tilde{r}_0^5-2097 \tilde{r}_0^2+\Bigg(2136 \left(A-3\right)+  5243 C\Bigg) \tilde{r}_0\\
\notag && +1424 C\left(3-A\right)\Bigg) \tilde{r}_0 +\tilde{q}^{12} \Bigg(356 \left(A-3\right)+   1221 C-4307 \tilde{r}_0\Bigg)+4 \left(D\right) \tilde{q}^3 \Bigg(\tilde{q}^4+2 \bigg(\tilde{B} \tilde{r}_0^5+3 \tilde{r}_0-  2 C\bigg) \tilde{r}_0 \tilde{q}^2\\
\notag && +\left(\tilde{B}^2 \tilde{r}_0^9+6 \tilde{B} \tilde{r}_0^5-4 \tilde{B} C \tilde{r}_0^4+\tilde{r}_0-4 C\right) \tilde{r}_0^3\Bigg) \Bigg(-42 \tilde{B}^3 \tilde{r}_0^{18}+534 \tilde{B}^2 \tilde{q}^2 \tilde{r}_0^{12}    +356 \tilde{B} \tilde{q}^4 \tilde{r}_0^6+89 \tilde{q}^6\Bigg) \Bigg) \\
\notag && - \frac{ 4 \Phi ^3\left(\tilde{r}_0-C\right)^3 }{531441 \tilde{r}_0^{21} \left(\tilde{B} \tilde{r}_0^6+\tilde{q}^2\right)^5} 
 \Bigg( 144 \tilde{B}^5 \Bigg(-5 \tilde{B}^2 \tilde{r}_0^9+21 \tilde{B} \tilde{r}_0^5+3 \tilde{B} C \tilde{r}_0^4+12 \tilde{r}_0-  31 C\Bigg) \tilde{r}_0^{35}\\
\notag && +8 \tilde{B}^4 \tilde{q}^2 \Bigg(15 \tilde{B}^3 \tilde{r}_0^{14}-59047 \tilde{B}^2 \tilde{r}_0^{10}+\tilde{B}^2 \left(13001 C-26012 \left(A-3\right)\right) \tilde{r}_0^9    +5548 \tilde{B}^2 C\left(A-3\right) \tilde{r}_0^8-91670 \tilde{B} \tilde{r}_0^6\\
\notag && +\tilde{B} \Bigg(105059 C-   45112 \left(A-3\right)\Bigg) \tilde{r}_0^5+48568 \tilde{B} C\left(A-3\right) \tilde{r}_0^4-6560 \tilde{r}_0^2-    3820 \left(A-3\right) \tilde{r}_0+39454 C \tilde{r}_0\\
\notag &&-20828 C\left(3-A\right)\Bigg) \tilde{r}_0^{28}-  4 \tilde{B}^3 \tilde{q}^4 \Bigg(9240 \tilde{B}^3 \tilde{r}_0^{14}+555171 \tilde{B}^2 \tilde{r}_0^{10}-3 \tilde{B}^2 \left(48895 C-74248 \left(A-3\right)\right) \tilde{r}_0^9  \\
\notag && -   46968 \tilde{B}^2 C\left(A-3\right) \tilde{r}_0^8+629759 \tilde{B} \tilde{r}_0^6+\tilde{B} \Bigg(306880 \left(A-3\right)-    728657 C\Bigg) \tilde{r}_0^5-324160 \tilde{B} C\left(A-3\right) \tilde{r}_0^4\\
\notag && +34864 \tilde{r}_0^2- 
  4 \left(49013 C-4898 \left(A-3\right)\right) \tilde{r}_0+103144 C\left(3-A\right)\Bigg) \tilde{r}_0^{22} \\
\notag && +\tilde{B}^2 \tilde{q}^6 \Bigg(-127971 \tilde{B}^3 \tilde{r}_0^{14}-4043860 \tilde{B}^2 \tilde{r}_0^{10}+\tilde{B}^2 \left(1111685 C-1668896 \left(A-3\right)\right) \tilde{r}_0^9+  \\
\notag && 345760 \tilde{B}^2 C\left(A-3\right) \tilde{r}_0^8-3616125 \tilde{B} \tilde{r}_0^6+24 \tilde{B} \Bigg(165291 C-  77944 \left(A-3\right)\Bigg) \tilde{r}_0^5+1921344 \tilde{B} C\left(A-3\right) \tilde{r}_0^4 \\
\notag && -179840 \tilde{r}_0^2+  \left(903799 C-98560 \left(A-3\right)\right) \tilde{r}_0-497408 C\left(3-A\right)\Bigg) \tilde{r}_0^{16}   -2 \tilde{B} \tilde{q}^8 \Bigg(81087 \tilde{B}^3 \tilde{r}_0^{14}+1852329 \tilde{B}^2 \tilde{r}_0^{10}\\
\notag && -12 \tilde{B}^2 \left(39905 C-71884 \left(A-3\right)\right) \tilde{r}_0^9-   175056 \tilde{B}^2 C\left(A-3\right) \tilde{r}_0^8+1336251 \tilde{B} \tilde{r}_0^6 \\
\notag && -6 \tilde{B} \Bigg(225213 C- 123200 \left(A-3\right)\Bigg) \tilde{r}_0^5-739200 \tilde{B} C\left(A-3\right) \tilde{r}_0^4+44960 \tilde{r}_0^2+  \\
\notag &&  \left(24640 \left(A-3\right)-217279 C\right) \tilde{r}_0+120896 C\left(3-A\right)\Bigg) \tilde{r}_0^{10}  \\
\notag &&  +\tilde{q}^{10} \Bigg(-86871 \tilde{B}^3 \tilde{r}_0^{14}-1845264 \tilde{B}^2 \tilde{r}_0^{10}+3 \tilde{B}^2 \left(135137 C-335744 \left(A-3\right)\right) \tilde{r}_0^9+  \\
\notag &&  200064 \tilde{B}^2 C\left(A-3\right) \tilde{r}_0^8-986814 \tilde{B} \tilde{r}_0^6+6 \tilde{B} \Bigg(155719 C-  93952 \left(A-3\right)\Bigg) \tilde{r}_0^5+549888 \tilde{B} C\left(A-3\right) \tilde{r}_0^4\\
\notag &&-17984 \tilde{r}_0^2-  9856 \left(A-3\right) \tilde{r}_0+82636 C \tilde{r}_0-46976 C\left(3-A\right)\Bigg) \tilde{r}_0^4+    \tilde{q}^{12} \Bigg(-10359 \tilde{B}^2 \tilde{r}_0^{10}\\
\notag && -477790 \tilde{B} \tilde{r}_0^6+2 \tilde{B} \left(37375 C-154048 \left(A-3\right)\right) \tilde{r}_0^5+  59776 \tilde{B} C\left(A-3\right) \tilde{r}_0^4-147785 \tilde{r}_0^2\\
\notag && +2 \Bigg(62077 C-44672 \bigg(A -3\bigg)\Bigg) \tilde{r}_0-84736 C\left(3-A\right)\Bigg) \tilde{r}_0^2+3225 \tilde{q}^{16}+32  \tilde{q}^3 D \Bigg(\tilde{B}^4 \Bigg(-6503 \tilde{B}^2 \tilde{r}_0^9 \\
\notag && +1387 \tilde{B}^2 C \tilde{r}_0^8-11278 \tilde{B} \tilde{r}_0^5+12142 \tilde{B} C \tilde{r}_0^4-955 \tilde{r}_0  +5207 C\Bigg) \tilde{r}_0^{28}+\tilde{B}^3 \tilde{q}^2 \Bigg(-27843 \tilde{B}^2 \tilde{r}_0^9+5871 \tilde{B}^2 C \tilde{r}_0^8\\
\notag && -38360 \tilde{B} \tilde{r}_0^5+40520 \tilde{B} C \tilde{r}_0^4  -2449 \tilde{r}_0+12893 C\Bigg) \tilde{r}_0^{22}+\tilde{B}^2 \tilde{q}^4 \Bigg(-52153 \tilde{B}^2 \tilde{r}_0^9+10805 \tilde{B}^2 C \tilde{r}_0^8-58458 \tilde{B} \tilde{r}_0^5+ \\
\notag && 60042 \tilde{B} C \tilde{r}_0^4-3080 \tilde{r}_0+15544 C\Bigg) \tilde{r}_0^{16}+\tilde{B} \tilde{q}^6 \Bigg(-53913 \tilde{B}^2 \tilde{r}_0^9+10941 \tilde{B}^2 C \tilde{r}_0^8 -  46200 \tilde{B} \tilde{r}_0^5+46200 \tilde{B} C \tilde{r}_0^4 \\
\notag && -1540 \tilde{r}_0+7556 C\Bigg) \tilde{r}_0^{10}+4 \tilde{q}^8 \Bigg(-7869 \tilde{B}^2 \tilde{r}_0^9+    1563 \tilde{B}^2 C \tilde{r}_0^8-4404 \tilde{B} \tilde{r}_0^5+4296 \tilde{B} C \tilde{r}_0^4-77 \tilde{r}_0+367 C\Bigg) \tilde{r}_0^4\\
\notag && +4 \tilde{q}^{10} \Bigg(-2407 \tilde{B} \tilde{r}_0^5+  467 \tilde{B} C \tilde{r}_0^4-698 \tilde{r}_0+662 C\Bigg) \tilde{r}_0^2+4 \tilde{q}^{12} \left(59 C-313 \tilde{r}_0\right)\Bigg)-2 \tilde{q}^{14} \Bigg(-4227 \tilde{B} \tilde{r}_0^6+ \\
\notag && 21176 \tilde{r}_0^2+20032 \left(A-3\right) \tilde{r}_0+947 C \tilde{r}_0+3776 C\left(3-A\right)\Bigg)    \Bigg)   \\
\notag && -\frac{64 \Phi ^5 \left(\tilde{r}_0-C\right) \left(\tilde{r}_0 \left(\tilde{r}_0 \left(3 \tilde{m}^2 \tilde{r}_0 \left(\tilde{r}_0-C\right)+2 \tilde{B} \tilde{r}_0^4+2\right)-4 C\right)+2 \tilde{q}^2\right)}{59049 \tilde{r}_0^7} +   \\
\notag && 64 \Phi ^4 \frac{\left(\tilde{r}_0-C\right)^2}{59049 \tilde{r}_0^{14} \left(\tilde{B} \tilde{r}_0^6+\tilde{q}^2\right)^2}  \Bigg( 2 \tilde{B}^2 \Bigg(2 \tilde{B}^2 \tilde{r}_0^{10}+6 \tilde{B} \tilde{m}^2 \tilde{r}_0^8-3 \tilde{B} \tilde{m}^2 C \tilde{r}_0^7+\tilde{B} \left(6 \tilde{r}_2-4 C\right) \tilde{r}_0^5-   3 \tilde{m}^2 C \tilde{r}_0^3-2 \tilde{r}_0^2 \\
\notag && +\left(4 C+6 \tilde{r}_2\right) \tilde{r}_0-12 C \tilde{r}_2\Bigg) \tilde{r}_0^{14}+\tilde{B} \tilde{q}^2 \Bigg(16 \tilde{B}^2 \tilde{r}_0^{10}+18 \tilde{B} \tilde{m}^2 \tilde{r}_0^8-  9 \tilde{B} \tilde{m}^2 C \tilde{r}_0^7-48 \tilde{B} \tilde{r}_0^6
\\
\notag && -8 \tilde{B} \left(3 A+C-18 \tilde{r}_2-9\right) \tilde{r}_0^5-2 \tilde{B} \Bigg(27 \tilde{r}_2 C+  12 C-4 AC\Bigg) \tilde{r}_0^4-3 \tilde{m}^2 C \tilde{r}_0^3-24 \tilde{r}_0^2-8 \Bigg(A-8 C- \\
\notag && 3 \tilde{r}_2-3\Bigg) \tilde{r}_0-6 \left(17 \tilde{r}_2 C+12 C-4 AC\right)\Bigg) \tilde{r}_0^8+\tilde{q}^4 \Bigg(15 \tilde{B}^2 \tilde{r}_0^{10}-    99 \tilde{B} \tilde{r}_0^6+9 \tilde{B} \left(-4 A+3 C+22 \tilde{r}_2+12\right) \tilde{r}_0^5 \\
\notag && -3 \tilde{B} \Bigg(27 \tilde{r}_2 C+12 C-  4 AC\Bigg) \tilde{r}_0^4+3 \tilde{m}^2 C \tilde{r}_0^3-12 \tilde{r}_0^2+\left(-4 A+41 C+12 \tilde{r}_2+12\right) \tilde{r}_0  -36 C+12 AC\\
\notag && -51 C \tilde{r}_2\Bigg) \tilde{r}_0^2-5 \tilde{q}^8+4  \tilde{q}^3 D \left(\left(C-3 \tilde{r}_0\right) \tilde{q}^2+\left(-3 \tilde{B} \tilde{r}_0^5+\tilde{B} C \tilde{r}_0^4-\tilde{r}_0+3 C\right) \tilde{r}_0^2\right) \bigg(2 \tilde{B} \tilde{r}_0^6  +\tilde{q}^2\bigg)\\
\notag && +\tilde{q}^6 \Bigg(-2 \tilde{B} \tilde{r}_0^6-6 \tilde{m}^2 \tilde{r}_0^4+3 \tilde{m}^2 C \tilde{r}_0^3-51 \tilde{r}_0^2+\left(-12 A+27 C+66 \tilde{r}_2+36\right) \tilde{r}_0  +4 C\left(A-3\right)-27 C \tilde{r}_2\Bigg) \Bigg) \Bigg) \,,
\end{eqnarray}
\normalsize{where}
\footnotesize
\begin{eqnarray}
\notag \Phi &=& \tilde{B} \left(-\frac{3 \tilde{r}_0}{C}-2\right)+\frac{\left(3 \tilde{q}^2+7 \tilde{r}_0^2\right) C}{\tilde{r}_0^7}-\frac{2 \left(4 \tilde{q}^2+\tilde{r}_0^2\right)}{\tilde{r}_0^6} \\
\notag \Upsilon &=& -12 \tilde{B}^2 \tilde{r}_0^{14} \left(-2 C+\tilde{B} \tilde{r}_0^5+\tilde{r}_0\right)+2 \tilde{B} \tilde{q}^2 \tilde{r}_0^7 \Bigg(\tilde{B} \tilde{r}_0^5 \left(27 C+58 \left(A-3\right)\right)+ 
 \tilde{r}_0 \left(58 \left(A-3\right)-181 C\right)\\
\notag && -116 C\left(A-3\right)+44 \tilde{B} \tilde{r}_0^6+ 
104 \tilde{r}_0^2\Bigg)+\tilde{q}^4 \tilde{r}_0 \Bigg(3 \tilde{B} \tilde{r}_0^5 \left(27 C+58 \left(A-3\right)\right)+\tilde{r}_0 \left(58 \left(A-3\right)-181 C\right) \\
\notag && -116 C\left(A-3\right)+150 \tilde{B} \tilde{r}_0^6+104 \tilde{r}_0^2\Bigg)+\tilde{q}^6 \Bigg(27 C
 +58 \bigg(A  
  -3\bigg)+50 \tilde{r}_0\Bigg)\\
  \notag && +58 \tilde{q}^3 D \left(\tilde{r}_0 \left(-2 C+\tilde{B} \tilde{r}_0^5+\tilde{r}_0\right)+\tilde{q}^2\right) \left(2 \tilde{B} \tilde{r}_0^6+\tilde{q}^2\right)  \\
\notag \Xi &=&-48 \tilde{B}^3 \left(\tilde{B} \tilde{r}_0^5+\tilde{r}_0-2 C\right) \tilde{r}_0^{21}+12 \tilde{B}^2 \tilde{q}^2 \Bigg(5 \tilde{B}^2 \tilde{r}_0^{10}-247 \tilde{B} \tilde{r}_0^6+\tilde{B} \Bigg(-111 A+71 C+  333\Bigg) \tilde{r}_0^5\\
\notag && +37 \tilde{B} C\left(A-3\right) \tilde{r}_0^4-86 \tilde{r}_0^2+\Bigg(-37 A+253 C+ 
 111\Bigg) \tilde{r}_0+111 C\left(A-3\right)\Bigg) \tilde{r}_0^{14}-3 \tilde{B} \tilde{q}^4 \Bigg(\tilde{B}^2 \tilde{r}_0^{10}+2151 \tilde{B} \tilde{r}_0^6+  \\
\notag && \tilde{B} \left(888 \left(A-3\right)-763 C\right) \tilde{r}_0^5-296 \tilde{B} C\left(A-3\right) \tilde{r}_0^4+344 \tilde{r}_0^2 
 +\left(148 \left(A-3\right)-1085 C\right) \tilde{r}_0+444 C\left(3-A\right)\Bigg) \tilde{r}_0^8+  \\
\notag &&  \tilde{q}^6 \Bigg(-288 \tilde{B}^2 \tilde{r}_0^{10}-4851 \tilde{B} \tilde{r}_0^6+\tilde{B} \left(2075 C-1776 \left(A-3\right)\right) \tilde{r}_0^5+ 
 592 \tilde{B} C\left(A-3\right) \tilde{r}_0^4-344 \tilde{r}_0^2\\
\notag &&  +2 \left(-74 A+561 C+222\right) \tilde{r}_0+ 
 444 C\left(A-3\right)\Bigg) \tilde{r}_0^2-102 \tilde{q}^{10}+\tilde{q}^8 \Bigg(-327 \tilde{B} \tilde{r}_0^6-1314 \tilde{r}_0^2+\Bigg(-444 A+ \\
\notag && 638 C+1332\Bigg) \tilde{r}_0+148 C\left(A-3\right)\Bigg)+148  \tilde{q}^3 D\left(\left(C-3 \tilde{r}_0\right) \tilde{q}^2+\left(-3 \tilde{B} \tilde{r}_0^5+\tilde{B} C \tilde{r}_0^4-\tilde{r}_0+3 C\right) \tilde{r}_0^2\right) \Bigg(3 \tilde{B}^2 \tilde{r}_0^{12} \\
\notag && +3 \tilde{B} \tilde{q}^2 \tilde{r}_0^6+\tilde{q}^4\Bigg)
\end{eqnarray}

\normalsize

\acknowledgments

We thank Kyriakos Destounis for carefully reading the manuscript and for his comments and suggestions. This work is partially supported by ANID Chile through FONDECYT Grant Nº 1220871  (P.A.G., and Y. V.). P. A. G. acknowledges the hospitality of the Universidad de La Serena where part of this work was undertaken.

\end{document}